\documentclass[sigconf]{acmart}
\usepackage{lineno,hyperref}

\usepackage{amsmath,amssymb,amsfonts,chemarrow,balance}
\usepackage{amsmath} 
\usepackage{graphicx}
\usepackage{subfigure}
\usepackage{mathenv}
\usepackage{color}
\usepackage{breqn}
 \usepackage{algpseudocode}
\newcommand{\BfPara}[1]{{\noindent\bf#1.}\xspace}
\usepackage{multirow}
\usepackage{float}
\usepackage{threeparttable}
\usepackage[inline]{enumitem}
\usepackage{wasysym}
\usepackage{etoolbox}

\usepackage{url}

\usepackage[T1]{fontenc}
\usepackage[utf8]{inputenc}
\usepackage{environ}
\usepackage{tikz}
\usetikzlibrary{arrows}
\usetikzlibrary{calc,matrix}
{\bfseries}{\itshape}
{\bfseries}{\itshape}
{\bfseries}{\itshape}
{\bfseries}{\itshape}
{\bfseries}{\itshape}
{\itshape}

\usepackage{xspace}
\usepackage{setspace}
\newcommand{\note}[1]{}

\newcommand{\bl}{{\em BlockAudit}\xspace}

\newcommand{\etc}{{etc.}\xspace}
\newcommand{\eg}{{\em e.g.}\xspace}
\newcommand{\ie}{{\em i.e.,}\xspace}

\usepackage[inline]{enumitem}

\newcommand{\etal}{{\em et al.}\xspace}

\definecolor{xgreen}{rgb}{0.2,0.6,0.0}
\definecolor{xred}{rgb}{0.7,0.1,0.0}
\def\equationautorefname~#1\null{(#1)\null}

\usepackage{mathtools}  
\usepackage{amsmath}
\usepackage{amssymb}
\usepackage{tabulary}
\usepackage{amsmath}

\usepackage[linesnumbered,ruled,noend]{algorithm2e}
\usepackage{algpseudocode}
\usepackage{amsfonts}
\usepackage{booktabs}

\definecolor{light-gray}{gray}{0.95}
\definecolor{darkgray}{rgb}{0.4, 0.4, 0.4}
\definecolor{purple}{rgb}{0.65, 0.12, 0.82}
\definecolor{editorGray}{rgb}{0.95, 0.95, 0.95}
\definecolor{editorOcher}{rgb}{1, 0.5, 0} 
\definecolor{editorGreen}{rgb}{0, 0.5, 0} 
\definecolor{orange}{rgb}{1,0.45,0.13}      
\definecolor{olive}{rgb}{0.17,0.59,0.20}
\definecolor{brown}{rgb}{0.69,0.31,0.31}
\definecolor{purple}{rgb}{0.38,0.18,0.81}
\definecolor{lightblue}{rgb}{0.1,0.57,0.7}
\definecolor{lightred}{rgb}{1,0.4,0.5}
\usepackage{upquote}
\usepackage{listings}

\newcommand\JSONnumbervaluestyle{\color{blue}}
\newcommand\JSONstringvaluestyle{\color{red}}

\newif\ifcolonfoundonthisline

\makeatletter
\lstdefinestyle{json}
{  
    backgroundcolor=\color{light-gray},
    basicstyle=\footnotesize\ttfamily,
    showstringspaces=false,
    breaklines=true,
    frame=lines,
  showstringspaces    =false,
  keywords            = {false,true},
  commentstyle= \itshape\color{codegreen},
  alsoletter          =0123456789.,
  morestring          = [s]{"}{"},
  stringstyle         = \ifcolonfoundonthisline\JSONstringvaluestyle\fi,
  MoreSelectCharTable =%
    \lst@DefSaveDef{`:}\colon@json{\processColon@json},
  keywordstyle        = \ttfamily\bfseries,
}
\newcommand\processColon@json{%
  \colon@json%
  \ifnum\lst@mode=\lst@Pmode%
    \global\colonfoundonthislinetrue%
  \fi
}

\lst@AddToHook{Output}{%
  \ifcolonfoundonthisline%
    \ifnum\lst@mode=\lst@Pmode%
      \def\lst@thestyle{\JSONnumbervaluestyle}%
    \fi
  \fi
  \lsthk@DetectKeywords%
}
\lst@AddToHook{EOL}%
  {\global\colonfoundonthislinefalse}
\makeatother

\acmDOI{10.475/123_4}

\acmISBN{123-4567-24-567/08/06}

\acmConference[ArXiv]{}
\copyrightyear{2019}

\acmArticle{4}
\acmPrice{15.00}

\begin{document}

\title{Secure and Transparent Audit Logs with BlockAudit}

\author{Ashar Ahmad}
\affiliation{%
 \institution{University of Central Florida}
}
\email{ashar@cs.ucf.edu}

\author{Muhammad Saad}
\affiliation{%
 \institution{University of Central Florida}
}
\email{saad.ucf@Knights.ucf.edu}

\author{Aziz Mohaisen}
\affiliation{%
 \institution{University of Central Florida}
}
\email{mohaisen@cs.ucf.edu}

\begin{abstract}
Audit logs serve as a critical component in enterprise business systems and are used for auditing, storing, and tracking changes made to the data. However, audit logs are vulnerable to a series of attacks enabling adversaries to tamper data and corresponding audit logs without getting detected. Among them, two well-known attacks are ``the physical access attack,'' which exploits root privileges, and ``the remote vulnerability attack,'' which compromises known vulnerabilities in database systems. In this paper, we present \bl: a scalable and tamper-proof system that leverages the design properties of audit logs and security guarantees of blockchain to enable secure and trustworthy audit logs. Towards that, we construct the design schema of \bl and outline its functional and operational procedures. We implement our design on a custom-built Practical Byzantine Fault Tolerance (PBFT) blockchain system and evaluate the performance in terms of latency, network size, payload size, and transaction rate. Our results show that conventional audit logs can seamlessly transition into \bl to achieve higher security and defend against the known attacks on audit logs. 
\end{abstract}

\maketitle

\section{Introduction}\label{sec:introduction}
Enterprise business systems and corporate organizations maintain audit logs for transparent auditing and provenance assurance~\cite{Wee99,OlivierS99}. In addition to their functional utility, the maintenance of audit logs is mandated by Federal laws. For instance, the Code of Federal Regulations of FDA, Health Insurance Portability and Accountability Act, \etc require organizations to maintain audit logs for data auditing, insurance and compliance~\cite{RingelsteinS09}. 

Secure audit logs enable stakeholders to audit the systems' state , monitor users' activity, and ensure user accountability with respect to their role and performance. Due to such properties, audit logs are used by data-sensitive systems for logging activities on a terminal database. Often times, audit logs are also used to restore data to a prior state after encountering unwanted modifications. These modifications may result from attacks by malicious parties, software malfunctioning, or simply user negligence. 

Audit logs typically use conventional databases as their medium for record keeping. Therefore, with databases, audit logs reflect a client-server model of communication and data exchange. The client-server model positions databases as a single point-of-trust for the audit logs, and therefore naturally a single point-of-failure. With this vantage of vulnerability, audit logs can be compromised in many ways. An adversary with root access to the database can manipulate critical information both in the database and the corresponding audit log. Once an audit log is compromised, the safety and transparency of the application is put to a risk. In the light of this weak security model, there is a need for secure, replicated, and tamper-proof audit logs that do not suffer from this shortcoming and have effective defense capabilities to resist attacks. To that end, we envision that blockchain technology can naturally bridge the gap to nicely serve the security requirements for audit log management, including ensuring security, provenance and transparency~\cite{BaruffaldiS18,FanRWLY18}. 

Over recent years, blockchain has acquired significant attention due to its use in distributed systems~\cite{DanezisM16}. In peer-to-peer settings, blockchain is capable of augmenting trust over an immutable state of system events~\cite{MauriCD18}. The most prominent example of blockchain technology has been realized in Bitcoin~\cite{StifterJSZW18}; a peer-to-peer digital currency that enables secure transfer of digital assets without the need of a trusted intermediary. Since Bitcoin, the use of blockchain has become prevalent in various applications and industries including smart contracts~\cite{kosba2016hawk,BhargavanDFGGKK16}, communication systems~\cite{SharmaRP18}, health care~\cite{GuoSZZ18,Rakic18}, Internet of Things~\cite{JesusCAR18,MAKHDOOM2019251}, censorship resistance~\cite{HeilmanBG16}, and electronic voting~\cite{DagherMMM18,HardwickA18}. The potential of blockchains is fully utilized in an environment where, 1) entities belonging to the same organization have competing interests~\cite{EljazzarAKE18} and/or 2) there is a need for immutable data management whose security increases over time~\cite{ZhangJ18a,Mettler16}. Because audit log applications meet the aforementioned requirements, they can intuitively use blockchain properties for an added security of audit logs.

Applied to an audit log application, blockchain can replicate the information contained in audit logs over a set of peers, thereby providing them a consistent and tamper-proof view of the system~\cite{AhmadSBM18}. Blockchains use an append-only model secured by strong cryptography hash functions. The security of data in the ledger increases while the blockchain grows with time. Furthermore, a malicious party intending to compromise the system will have to change logs maintained by a majority of peers. This increases the cost and complexity of the attack and increasing the overall defense capability of the audit log application. However, the design space of blockchain is modular due to varying access control policies and consensus schemes. Therefore, it becomes a design challenge to apply suitable structural and functional primitives that best fit the application requirements and achieve the end goal of transparency and provenance. Motivated by that, we propose a blockchain-based audit log system called \bl. Broadly speaking, in \bl, we 1) capture system events generated by the data access layer of an enterprise application, 2) transform the acquired information into blockchain compatible transactions, 3) construct a peer-to-peer network consisting of entities that evaluate and approve the authenticity of transactions by executing a consensus protocol, and 4) lock the transaction in an append-only and immutable blockchain ledger, maintained by each network entity. 

\BfPara{Contributions}
In summary, in this paper we make the following key contributions: 
\begin{enumerate*}

    \item We outline security vulnerabilities in audit log applications and discuss shortcomings of the prior work in addressing those vulnerabilities. 

    \item We present a blockchain-based audit log system called \bl which addresses these vulnerabilities and ensures security, transparency, and provenance in the auditing system. Towards that, we review the modular constructs of blockchain systems and discuss suitable design choices that best fit the requirements of an auditing system. 
    
    \item We test the design of \bl using a real-world eGovernment application, provided by Clearvillage Inc, and analyze its performance using three evaluation metrics, namely the latency, the network size, and the payload size.  

    \item Based on observations made from theoretical analysis and experiments, we discuss our proposed solution and provide future directions for research on blockchain-based audit logs. 
 
\end{enumerate*}

\BfPara{Organization}
The rest of the paper is organized as follows. 
In \textsection\ref{sec:background}, we provide the background and the threat model. In \textsection\ref{sec:rw}, we review the notable work done towards audit log security. In \autoref{sec:problem}, we outline the problem statement and the design requirements based on the limitations of prior work. In \autoref{sec:design}, we present our proposed solution called \bl, followed by its analysis in \autoref{sec:Analysis}. Experiments and evaluations are presented in \autoref{sec:eval}, followed by a discussion and concluding remarks in \autoref{sec:discusion} and \autoref{sec:conclusion}, respectively. 

\section{Background and Threat Model} \label{sec:background}
In this section, we provide the background of audit logs including their benefits and vulnerabilities. We also provide a threat model for the systematic exposition of the outlined vulnerabilities.

\subsection{Audit Logs} \label{sec:auditlogs}
\begin{figure}[t]
\begin{center}
    \includegraphics[width=0.45\textwidth]{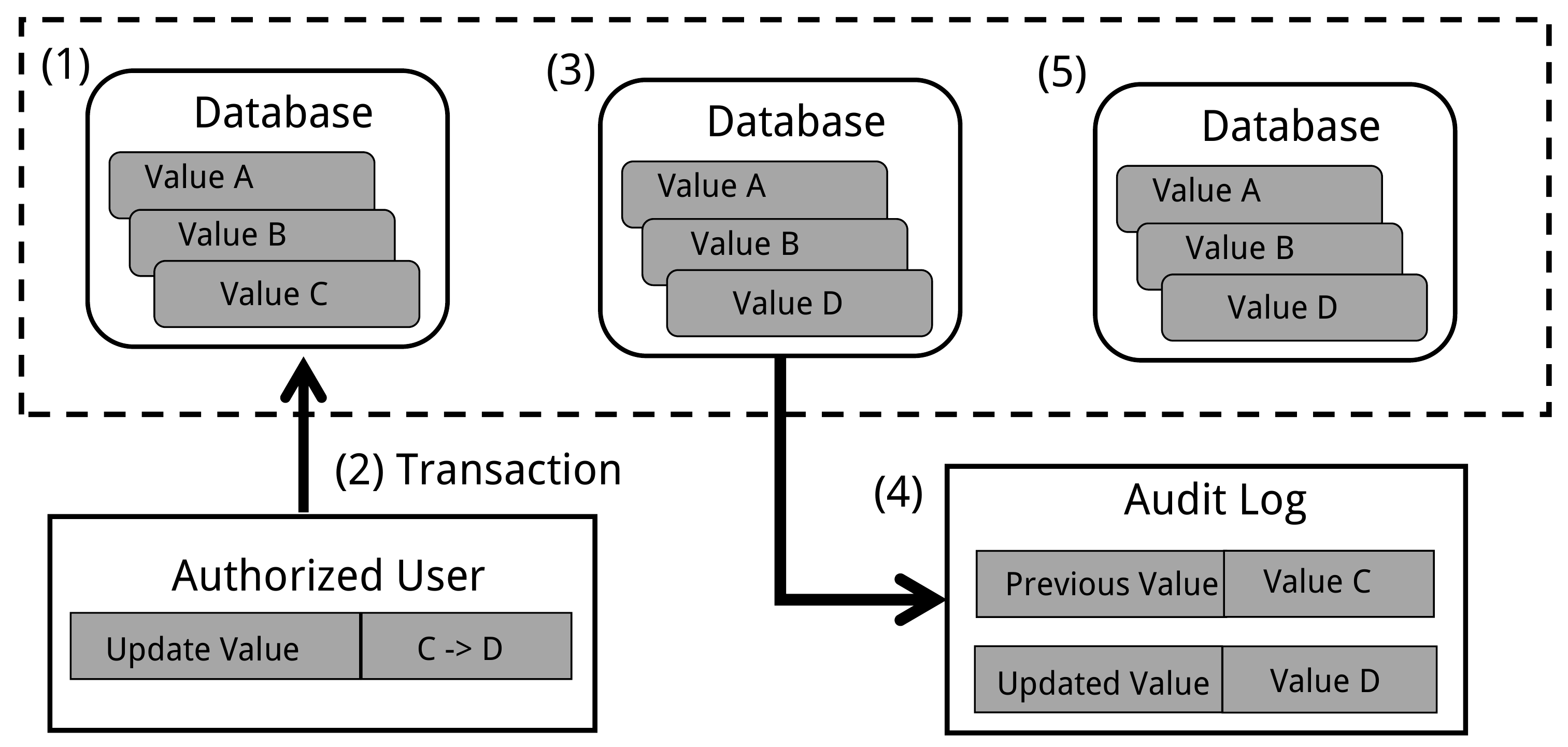}
\caption{Audit log generation in an OLTP system. We annotate each step with a number to show the sequence of progression. Notice that the user generates a transaction to change the value from C to D, and the change is then recorded in the audit log by the database.}
\label{fig:DesignOverview}
\end{center}
\end{figure}

An audit log is an essential component in {\em online transaction processing} (OLTP) systems such as order entry, retail sales, and financial transaction systems~\cite{VieiraM03,NikolaouMG97}. The OLTP system maintains audit logs to monitor users' activity and provide insight into the sequential processing of transactions~\cite{SirinYA17}. Each processed payment in OLTP system creates a unique record in the audit log. The aggregate volume of transactions and the total payment made during a financial year can be verified by consulting the data recorded in the audit logs. Moreover, these audit logs can also be used to identify discrepancies, anomalies, and malicious activities in payments. Audit logs have to be secure, searchable, and readily accessible from the application so that business users can easily view the chain of actions that lead to the current state of a business object. In \autoref{fig:DesignOverview}, we provide an overview of the OLTP system in which an audit log is generated once an authorized user commits a transaction to the database. The transaction makes a change in the value of an object and, as a result the change is recorded in the database and audit log. These changes can be matched later with the database and/or the application for auditing and provenance.

\BfPara{Benefits of Audit Logs} The audit log is widely used in modern information systems to provide a chronological record of changes being made to the data, and track the life-cycle of objects. Audit logs are also used to verify and authenticate operational actions, provide proof-of-compliance, ensure operational integrity, detect malicious activity, and provide system-wide provenance~\cite{ZhaoQZZL18,Hartung16}. Organizations that use audit log applications no longer maintain a paper trail for chronological record management, thereby saving cost and storage space with additional environmental benefits. With the elimination of the paper trail, the electronic audit logs are solely responsible for establishing security and the correctness of sensitive information. In the situation of an attack, the audit log is typically used as a starting point of forensic analysis.  

Audit logs also contribute to organizing user behavior in applications. Since audit logs maintain the user activity over time and detect misbehavior, naturally, they promote responsible user behavior and reduce the chances of misconduct. The users remain aware of their actions being recorded in an audit log. Moreover, in the case of an attack or a malicious activity, audit logs can be used to ensure users are accountable for their actions.

The correctness of audit logs is imperative to find the cause of the attack and initiate suitable countermeasures. For example, the first step in identifying the solution to a system crash is obtaining appropriate knowledge of the conditions that lead to the crash. This knowledge can be obtained through audit logs which can be used to reconstruct the conditions of an event. Such reconstruction can correctly identify the root cause of the issue, such as network failures, system bugs, or information tampering. Furthermore, after fault detection, the system can be restored back to the original state by rolling back transactions to the point in time prior to the attack. Atop the real-time monitoring, audit logs can also be used to identify system-related problems such as implementation errors, software bugs, and deployment faults. Finally, audit logs can also help in intrusion detection, by providing useful information to detect unauthorized system access.

\BfPara{Vulnerabilities in Audit Logs} 
Despite the aforementioned benefits, audit logs are vulnerable to a series of attacks that may compromise the integrity of OLTP systems. An attacker can use multiple attack vectors which exploit the known weaknesses in OLTP systems and corrupt the state of the database and audit logs. Conventional schemes of protecting audit data include the use of an append-only device such as continuous feed printer or {\em Write Once Read Multiple} (WORM) optical devices. These systems work under a weak security assumption that the logging site cannot be compromised, which eventually keeps the integrity of the system intact. However, attackers have often exploited vulnerabilities at logging site to tamper with data in audit logs~\cite{LeeZX13,Margulies15}.

If the attacker acquires the credentials of an authorized user, he can corrupt the database as well as the audit log. On the other hand, if the attacker compromises the database by breaching its defense, he can manipulate the database and prevent it from populating audit logs. Then, not only he will be able to corrupt the database, but also disable the auditing procedure by blocking the backward compatibility of audit logs with the database.   

\subsection{Threat Model}\label{sec:thm}
To sufficiently analyze the vulnerabilities of audit logs and set the security model objectives, we present the threat model for the auditing systems in this section. 

Inspired by the limitations found in the prior work~\cite{LeeZX13,Margulies15}, our threat model assumes an adversary that is capable of both physically accessing the trusted computing base (TCB) and remotely penetrating the OLTP system by exploiting software bugs.  As such, the adversary can be a malicious third party aiming to tamper data to compromise auditing procedures. This would require the adversary to obtain root privileges to the system, or have significant knowledge of the system architecture. Additionally, the adversary can also hack and acquire the credentials of a root user of the system. This can be carried out using various attack procedures available in the conventional attack catalog~\cite{LuhMKJS17}. However, possessing the knowledge of a private database system or a remotely acquiring credentials of a root user would require exceptional capabilities for the adversary. Therefore, we assume the third party attacker to have strong capabilities. 

In a less hostile environment, the adversary can also be someone from within the system with root privileges. For instance, a corrupt auditor, who has tampered data for personal gains, might want to cover his act by changing data values. In contrast to the third party attacker, this adversary will not need sophisticated capabilities since he already has root privileges and the system knowledge. 

For the system architecture, we assume an OLTP system similar to a retail sale repository. The system implements the design logic of an application using secure communication protocols such as SSL/TLS. Moreover, the system has a database that keeps records of sales and maintains a remote audit log. The audit log keeps track of the database changes through transactions, as shown in~\autoref{fig:DesignOverview}. In such a design, the attacker can exploit the system by launching two possible attacks, namely the physical access attack and the remote vulnerability attack. 

\BfPara{The Physical Access Attack} 
In the physical access attack, the adversary will use the root privileges to corrupt the database. As shown in \autoref{fig:DesignOverview}, the adversary will generate a series of transactions to change the values of objects in the database. Once the attacker manipulates the data, the database will automatically generate an audit log, tracking all changes made by the attacker. However, to evade detection, the attacker can either delete the newly generated audit log or modify its values. Furthermore, the attacker will also be able to tamper the history maintained by the audit log in order to corrupt the auditing process. Therefore, in the physical access attack, we assume an adversary inside or outside the system who has access to the key system components. 

\BfPara{The Remote Vulnerability Attack}
In the remote vulnerability attack, the attacker may only exploit the default vulnerabilities in the OLTP applications such as software malfunctions, malware attacks, buffer overflow attacks \etc. In this attack, the adversary, although not as strong as the physical access attack may still be able to contaminate the database and the audit log with wrong information. Despite these adversarial capabilities, we assume that the OLTP application is secure against the conventional database and network attacks such as SQL injection and weak authentication. Generally, database systems used by corporate organizations are secure against these conventional attacks, and for the application service used in this paper, we ensure this requirement is meet.

\begin{figure}[t]
\begin{center}
    \includegraphics[width=0.45\textwidth]{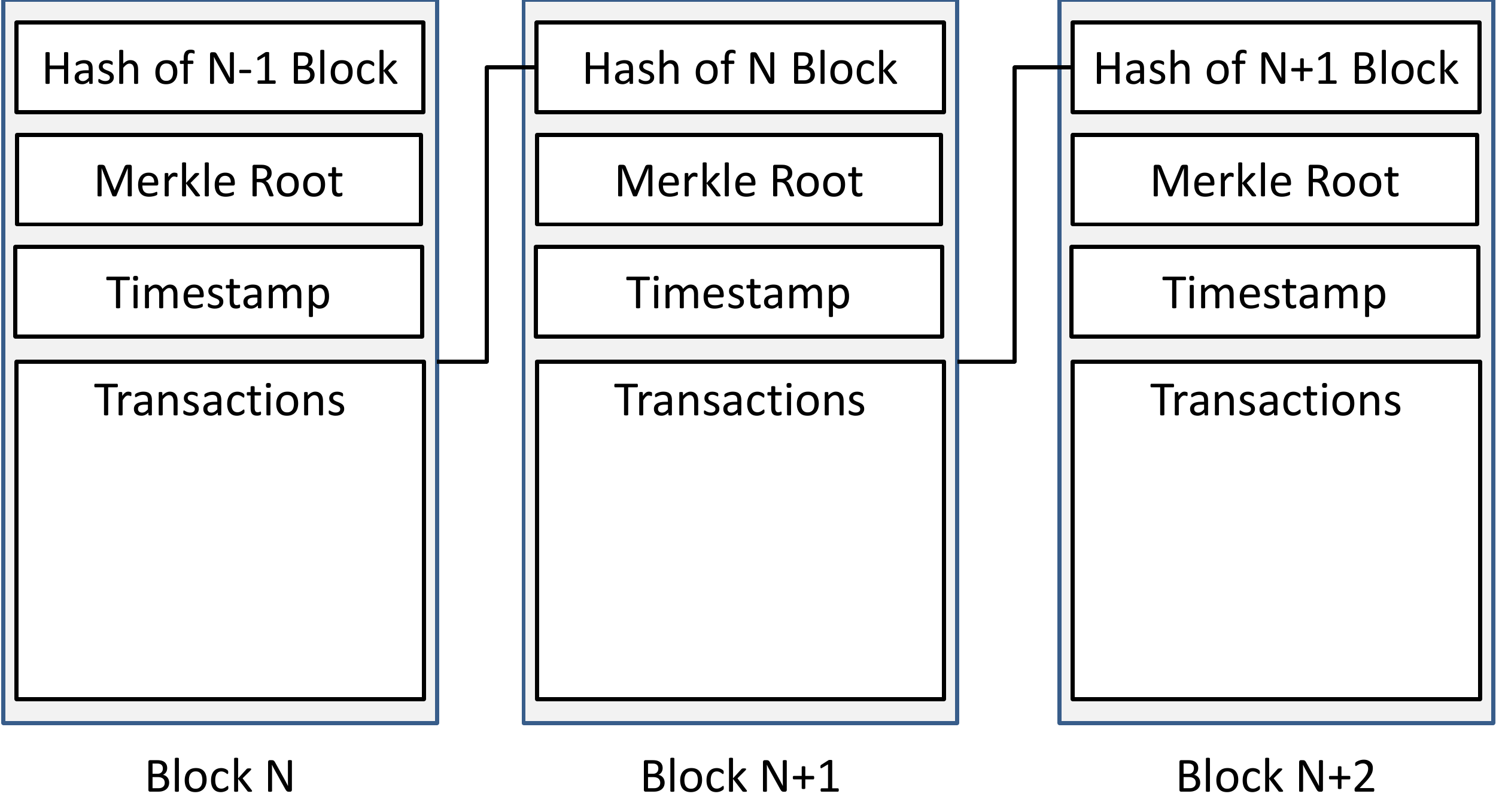}
\caption{An overview of Blockchain structure consisting of three blocks. Notice that each block header consists of the hash of the previous block. This relationship gives blockchain, the property of an immutable ledger. Also notice that the merkle root ensures that the transactions are ordered in a sequence. }\label{fig:blockchain}
\end{center}
\end{figure}

\section{Related Work} \label{sec:rw}
In the following, we review the notable work done in the direction of securing audit logging mechanisms. We also discuss the limitations of the prior work in light of the threat model (\textsection\ref{sec:thm}).

\BfPara{Audit Logs}
Schneier and Kelsey \cite{SchneierK99, SchneierK98} proposed a secure audit logging scheme capable of tamper detection even after compromise. However, their system requires the audit log entries to be generated prior to the attack. Moreover, their system does not provide an effective way to stop the attacker from deleting or appending audit records, which, in our case is easily spotted by \bl. Snodgrass \etal \cite{SnodgrassYC04} proposed a trusted notary based tampering detection mechanism for RDBMS audit logs. In their scheme, a check field is stored within each tuple, and when a tuple is modified, RDBMS obtains a timestamp and computes a hash of the new data along with the timestamp. The hash values are then sent as a digital document to the notarization service which replies with a unique notary ID. The ID is stored in the tuple, and if the attacker changes the data or the timestamp, the ID becomes inconsistent, which can be used for attack detection. Ray \etal \cite{ray2013secure} proposed a framework for maintaining secure audit logs in cloud computing platforms. In particular, their framework uses cryptography to maintain integrity and confidentiality while storing, processing, and accessing the audit logs. Ma and Tsudik \cite{MaT09} proposed a technique to generate an aggregate signature by sequentially combining individual log entry signatures using forward-secure, append-only signatures. This scheme provides provable security with efficient space utilization; where the correctness of individual entry can only be verified by generating the aggregated signature. Yavuz \etal \cite{YavuzNR12} proposed a scheme that stores individual and aggregate signatures, where the storage of individual signatures increases the storage footprint while allowing individual verification of signatures.

\BfPara{Blockchains} A blockchain is a data structure that enables transparent and tamper-proof data management in distributed systems \cite{HyvarinenRF17, HolotiukPM17}.  As such, blockchain consists of a sequence of data blocks that are linked through on-way hash functions. Due to the one-way property of hash operations, blockchain exhibit the append-only model where once a data item is inserted it becomes immutable \cite{HalunenVK18,DerlerSSS19}. An illustration of the blockchain data structure is provided in Figure \ref{fig:blockchain}. Transaction ordering using blockchain is enabled by multi-party consensus schemes \cite{ShahaabLHK19,YiYWJ19}. Popular among these schemes are the proof-of-work, proof-of-stake, and practical Byzantine Fault Tolerance \cite{WangDZ18, HalunenVK18}. Roughly speaking, a consensus algorithm is a set of instructions executed independently by each party in the system. The execution is completed if a majority under fixed bound obtains the same output from the computation. For more on blockchains and consensus schemes, we refer the reader to \cite{SaadCLTM19,SaadAAAYM19,BanoBAMMD17}.

\BfPara{Blockchain and Audit Logs} Combining blockchain and audit logs, Sutton and Samvi \cite{semwebSuttonS17} proposed a blockchain-based approach that stores the integrity proof digest to the Bitcoin blockchain. Bitcoin uses a proof-of-work (PoW) consensus protocol. As we show later in \autoref{tab:ca}, PoW suffers from low throughput and high confirmation time. In particular, Bitcoin has a maximum throughput of 3--7 transactions per second. Therefore, for audit log applications that have a high transaction generation rate, the concept provided in \cite{semwebSuttonS17} can be insufficient. Castaldo \etal \cite{Castaldo2018} proposed a logging system to facilitate the exchange of electronic health data across multiple countries in Europe. They created a centralized logging system that provides {\em traceability} through {\em unforgeable} log management using blockchain. Cucrull \etal \cite{CucurullP16} proposed a system that uses blockchain to enhance the security of the immutable logs. Log integrity proofs are published in the blockchain providing non-repudiation security properties.   

\section{Problem Statement \& Requirements} \label{sec:problem}
 The prior related research provides the groundwork for securing audit logs with blockchains and represent the foundation of our work. However, our major contribution is seen in our focus on audit logs related to enterprise business applications, focusing on scalability and performance. As outlined in \textsection\ref{sec:introduction}, blockchain applications may vary in their access control policies and consensus schemes. Exploring the blockchain model for Enterprise business applications would require an understanding of their requirements, and methods to overcome the domain-specific design challenges, which we explore in this paper.

Another limitation that can be observed in \cite{CucurullP16, SnodgrassYC04} is the inability to address Byzantine behavior among network peers. In other words, the application assumes all participating entities faithfully execute the consensus protocol without incurring any malicious behavior. However, in distributed systems adversaries can control a subset of replicas who can behave arbitrarily in order to withhold transaction processing and cause conflicting views among other replicas. Tolerance towards Byzantine nodes is a function of consensus schemes to be applied. For instance, permissionless blockchain applications such as Bitcoin can tolerate up to 50\% of Byzantine nodes while maintaining operational consistency. On the other hand, PBFT-based private blockchains can tolerate only 30\% Byzantine nodes. Therefore, the selection of a consensus algorithm can influence the security model of the application. In \bl, we address the aforementioned limitations and present an end-to-end solution constructed by transforming knowledge problems into design problems.

\subsection{Design Engineering}
So far, we have discussed the benefits of audit logs, their key vulnerabilities, and the existing solutions that address those vulnerabilities. We have also presented a threat model to outline adversarial conditions. In this section, we use this knowledge to make design choices to meet the requirements of a practical blockchain-based audit log solution. In the following, we define functional, structural, and security requirements that we expect \bl to meet. 

\subsubsection{Functional Requirements} \label{sec:funrq}
An audit log application is expected to ensure trust in the application data and provide tamper-proof evidence of transaction history when needed. Data tampering has to be prevented for the application data as well as the audit logs. However, a priority is given to the audit logs, since they are used to establish provenance. For this purpose, the audit log data should be stored across multiple peers in such a way that it remains consistent at each node, and therefore, hard to corrupt. If tampering happens at any node, the system should be able to detect and correct it. This requirement, however useful, comes with an assumption that a majority of peers behaves honestly, and faithfully executes the system protocols. 

For audit data to be added to the blockchain, the participating peers in the audit log network must reach a quick consensus over a newly generated transaction. Since audit logs are generated in real-time and persisted inside the database transaction, therefore, any delay in using distributed audit logs adversely affects the system performance. In order to prevent such delays, the system needs to have low latency while maintaining the capability of processing a large volume of transactions. Additionally, the application should not add any data without consensus among a majority of peers.

The audit log system architecture should be modular and service-oriented so that it is possible for various types of applications to participate and benefit from this system. Moreover, audit logs should be data agnostic and must not rely upon the nature of data that is stored in them. The business application should be able to provide data in any format as per the requirements of the application.

Finally, the audit log system should provide searching and retrieval capability to enable the retrieval of any desired transaction or a set of transactions (e.g., audit log entries for the last ten minutes, all audit log entries registered against a specific user ID, \etc). The search needs to be fast and responsive to ensure the end user is able to perform the audit in real-time. 

\begin{figure}[t]
\centering
\includegraphics[width=0.35\textwidth]{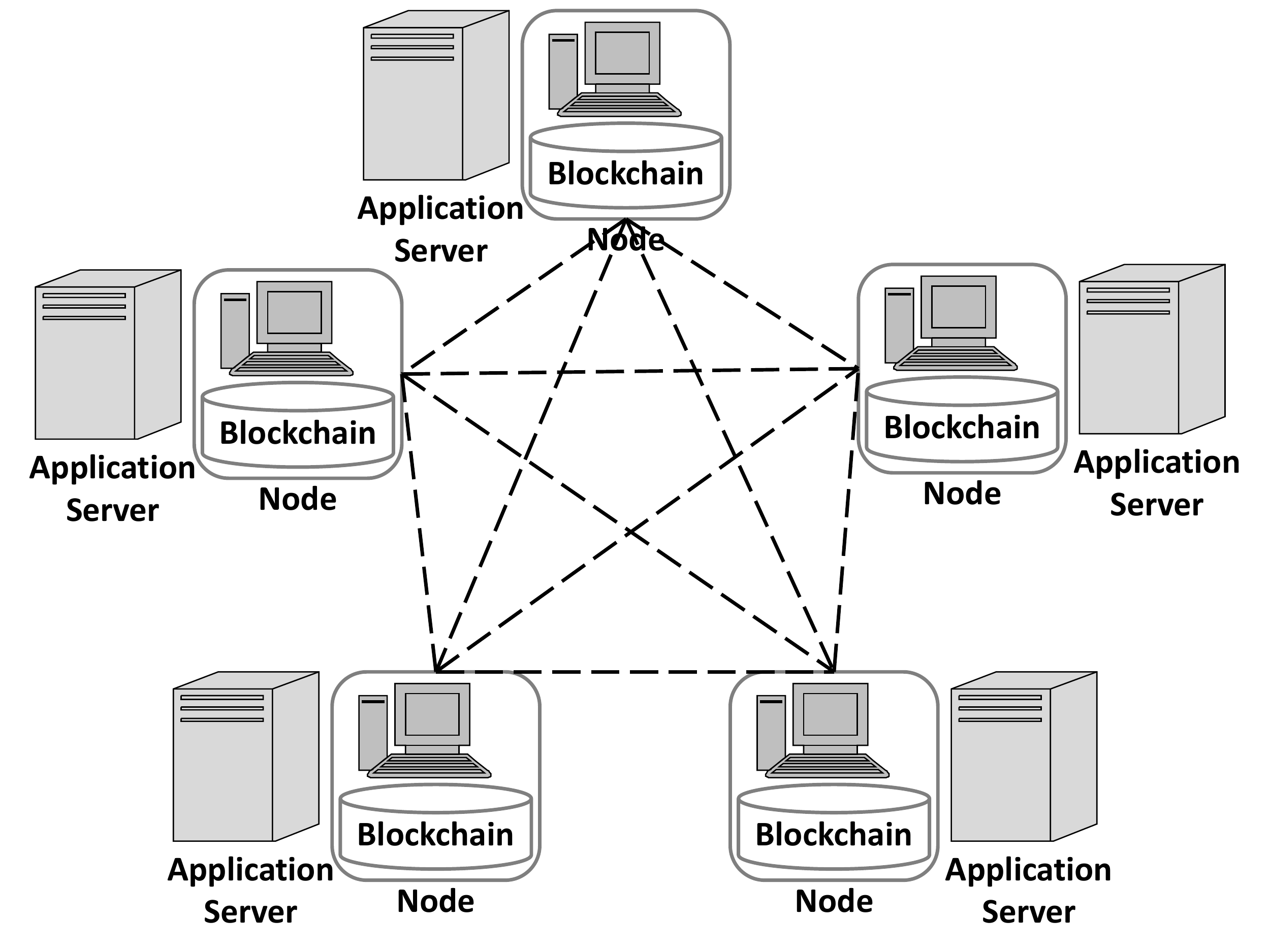}
\caption{The network overview of nodes employing \bl. Notice that each node maintains an interface that connects them to the audit log application. They exchange transactions with one another during the application life-cycle. }
\label{fig:server}
\end{figure}

\subsubsection{Structural Requirements} \label{sec:strucrq}
Keeping in view of the design the baseline models introduced  \cite{CucurullP16, SnodgrassYC04},  we envision that \bl must operate in a distributed manner with application services running on multiple hosts without a central authority.
As such each application peer would require its own blockchain node to become part of the the \bl network. The audit log system should have a high throughput and should be able to process a large number of transactions. \bl should be able to support transactions of various sizes since the transaction size varies in audit log applications. The audit log system should be easy to integrate with existing system with minimal structural and functional changes in the application. It should also be independent of the underlying application database. Finally, the system auditing should be secure, transparent, and visible to all peers within the network.  

\subsubsection{Security Requirements} \label{sec:secrq}
In the light of our threat model \textsection\ref{sec:thm}, we require \bl to be secure in adversarial conditions. To that end, if the adversary launches a physical access attack, \bl should be able to neutralize it and prevent data tampering at the source. If the adversary launches the remote vulnerability attack, \bl should stop the attack propagation across the network peers. In other words, if the adversary exploits a bug in the audit log of one peer, \bl should immediately recognize the attack and notify the victim peer. Furthermore, the infection should be curtailed at the target zone, preventing its spread in the network.

In addition to the baseline attack model, we also expect \bl to remain secure in the presence of Byzantine nodes. Therefore, if a strong adversary controls a subset of nodes in the network, he should not be able to corrupt audit logs or delay transaction verification. This can be achieved by either raising the attack cost \ie constructing a large network or relaxing anonymity so that the adversary risks identity exposure by misbehaving. 

\section{\bl}\label{sec:design}
In this section, we show the implementation of \bl. First, we describe the eGovernment application that we used to generate audit logs. Next, we show how a blockchain network is constructed to integrate audit logs. In that, we describe the methods of generating transactions, creating a distributed network, managing the access control, and developing consensus among peers over the state of the audit logs. 

\begin{figure}[t]
\begin{center}
    \includegraphics[width=0.48\textwidth]{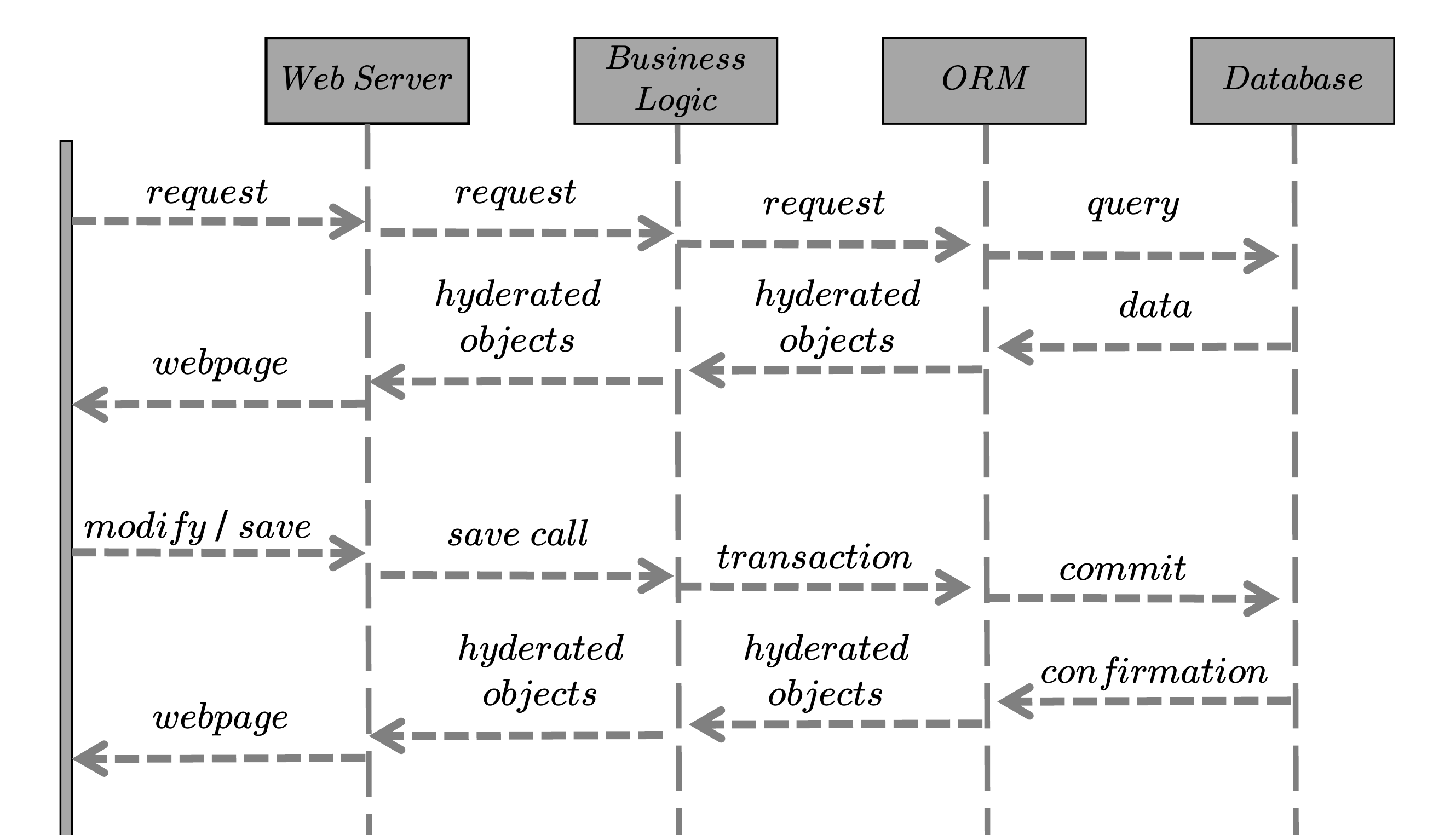}
\caption{The information flow between various components of the application. Notice that the transaction is generated at the business logic layer, and once the database commits to the transaction it is rendered on the web page.}
\label{fig:ormmodel}
\end{center}
\end{figure}

\subsection{Application Architecture} \label{sec:dch}
For \bl, we use an eGovernment application provided by a company called ClearVillage inc which provides software solutions for various government operations \eg property appraiser, building permits \etc. The application uses a multi-tier system architecture comprising of web and mobile clients, a business logic layer, a data access layer, and a database. In the following, we describe the core functionality of each component along with its role in generating an audit log.  

\BfPara{Web Applications} The web applications are built using {\em asp.net} and users access the application services through a web browser. Additionally, native clients are provided for Android and iOS, built using their respective development frameworks. The web application and web services are hosted on Microsoft's Internet Information Services (IIS) web server. The public side portal is available on the Internet and gives public users access to information without authentication. Atop this, a staff portal is provided to the organization staff which is only accessible from within the organization, thereby providing another security layer. 

\BfPara{Business Logic Layer} 
The business logic layer is an interface between clients and the database layer, responsible for implementing business rules. Among other functions, the business logic layer also manages data creation, data storage, and changes to the data with the help of object-relational mapping (ORM). Upon receiving a request from the client, the web server instantiates the relevant objects in the business logic layer, which uses the ORM to send the processed object to the client. The ORM writes changes to the objects in the RDBMS tables.

\BfPara{ORM} The ORM in the application provides a mapping mechanism that allows querying of data from RDBMS using an object-oriented paradigm~\cite{AlghamdiOC16, Bagheri0S17}. Modern web applications are well suited for this technique since they are multi-threaded and are rapidly evolving. ORM also reduces the code complexity and allows developers to focus on business logic instead of database interactions. This application uses {\em NHibernate}\cite{Nhibernate}: an ORM solution for Microsoft .NET platform. {\em NHibernate} is a framework used for mapping an object-oriented domain model to RDBMS and it maps the .NET classes to database tables. It also maps Common Language Runtime (CLR) data types to SQL data types. The ORM inside a database layer creates a SQL statement to hydrate the object and passes it to the business logic layer. ORM also flushes the changes to the RDBMS and commits a transaction. Interactions between the application and RDBMS are carried out using the ORM. In \autoref{fig:ormmodel}, we provide the information flow between application components.

\begin{figure}[t]
\begin{center}
\includegraphics[width=0.4\textwidth]{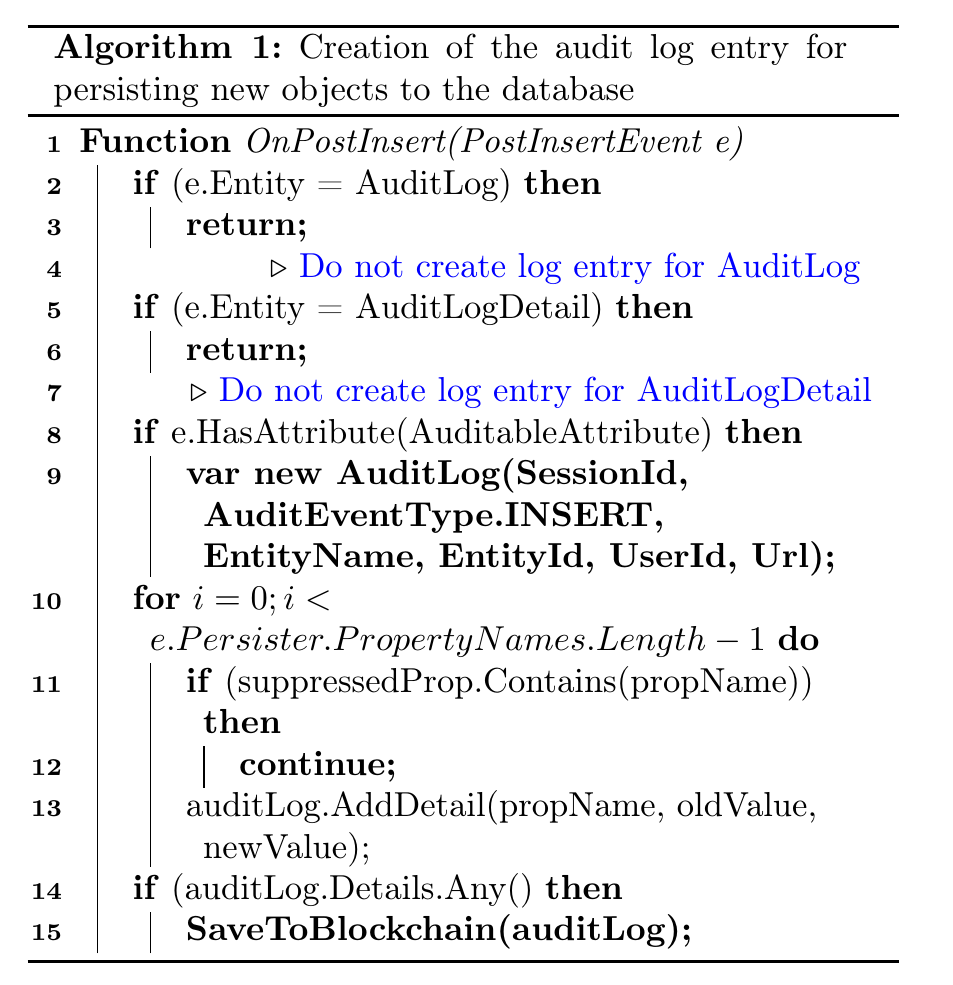}
\label{algo:AuditCreate}
\end{center}
\end{figure}

\begin{figure}[t]
\begin{center}
\includegraphics[width=0.4\textwidth]{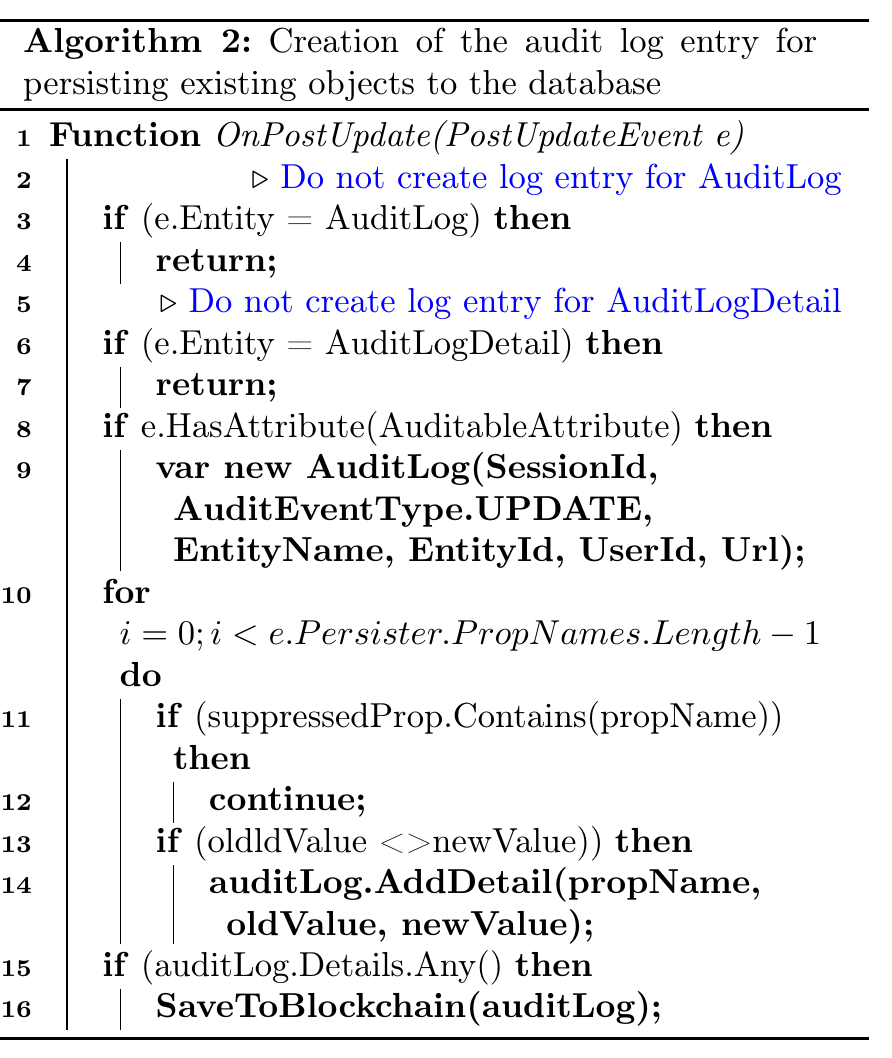}
\label{algo:AuditUpdate}
\end{center}
\end{figure}

\subsection{Generating Audit Logs}
In this section, we show how the application generates an audit log once the user commits a transaction. To implement auditing, three events provided by {\em nHibernate} are used, namely {\em IPostInsertEventListener}, {\em IPostUpdateEventListener}, and {\em IPostDeleteEventListener}.

{\em IPostInsertEventListener} event is triggered once a transient entity is persisted for the first time. Each class that requires auditing is marked with {\em Auditable} attribute, which is then used to create audit logs for classes containing this attribute. All mapped properties are then audited by default and a suppress audit attribute is added to suppress auditing of a target property. Usually, and by default, all properties are audited. However, in special cases where auditing is not required, the {\em SuppressAudit} attribute is added to the property. In Algorithm 1, we show the process of generating the audit log when {\em IPostInsertEventListener} event is triggered.  

When an audit entry is created, it contains a session ID (transaction ID), a class name, an event type (Insert, Update, or Delete), audit ID, creation date, user ID, URL, and a collection of values for all properties. The collection of values consists of the old value before the update and the new value resulting from the update. Moreover, during an update, old and new values are compared. Only if the two values are different from one another, the change is committed to the audit log. In Algorithm 2, we outline this procedure of updating audit logs. Currently, these audit logs are saved inside an RDBMS using two tables, the AuditLog table, and the AuditLogDetail table. Furthermore, Globally Unique Identifiers(GUID) are used as primary keys in auditlog tables.

Once a change is observed in a class, the ORM's event handler is invoked. Similarly, the event handler is also invoked when the change is observed in the ``AuditLog'' and the ``AuditLogDetail'' classes. Lines 2--5 in Algorithm 1 and Algorithm 2, prevent the creation of logs for Audit Classes. In the absence of this condition, the event logger would fall into an infinite event loop.The infinite loop can also be prevented by removing the ``AuditableAttribute'' from the audit classes. However, we use lines 2--5 as a check to avoid the loop in case a developer adds the attribute by mistake. 

Once an audit log is generated, the application provides a link to the audit log page from the primary object. The link allows end users to look at the object history and track any discrepancy caused by a bug or malicious activity.

\subsection{Blockchain Integration to Audit Logs} \label{sec:bcaudit}
In this section, we will show how audit logs, obtained from our application, are integrated with the blockchain. So far in our design, we have an application that stores audit logs upon receiving a transaction. Now, we need to convert the audit log data into a blockchain-compatible format (blockchain transactions) and construct a distributed peer-to-peer network to replicate the state of the blockchain over multiple nodes. In our current implementation the audit log is generated using the ORM, which calls a Representational State Transfer(REST) Application Programming Interface(API) to store the audit log entry. 

We used the ORM to create audit logs because the ORM acts as the gateway to capture all database transactions. Therefore, it is efficient to take advantage of ORM events to capture all the database changes and convert them into a JSON packet for the REST API. Our design is flexible and generic, and can also be used by other applications that do not use the ORM. Other than the ORM, the application layer or the data access layer can also be extended to capture the database changes in a JSON format and invoke the REST API. Moreover, the REST API can also be used by applications built using a serverless architecture.

\subsubsection{Creating Blockchain Network} \label{sec:cbn}
In \bl, the network consists of peers that all have the privilege of accessing the application and creating an audit log. This network is connected in peer-to-peer model~\cite{GaneshKM03} and each peer can connect to all the other peers in the network. Connecting to a bigger subset of peers is beneficial, because it can avoid unnecessary delays in receiving critical information. 

\BfPara{Access Control} \label{sec:ac}
As mentioned in \autoref{sec:introduction}, access controls may vary across blockchain application. These applications can be permissionless (open access) or permissioned (selective access). In permissionless applications, such as Bitcoin, an arbitrary user can download the Bitcoin Core software and join the network. However, in the private and permissioned blockchains, an access control mechanism is applied that restricts the participation to only approved users. Since audit logs consist of sensitive data, therefore, in \bl we use a permissioned blockchain with access control provisioned to selected users. In permissioned blockchains, adjusting access control is trivial since any custom membership service can be used for the access control~\cite{Bano:2017b}. To avoid runtime complexities, we do peer screening prior to network creation. The peer screening is done based on the IP addresses in which we curate a list of IP addresses, compile them in executable code, and provide the code to each peer. Upon executing the code, the peer gets connected to the network. 

\begin{table}[t]
\centering
\caption{{\em Clear Village's} actual transaction sizes in (bytes) for the three transaction schemes, based on the transactions from October 2018. The average size is between 32 bytes to 9,302 bytes.} \label{tab:transations}
\scalebox{0.9}{
\begin{tabular} {l|c|c|c}
\hline
\textbf{Type} & \textbf{Max} & \textbf{Min}& \textbf{Average} \\ \hline
\textbf{Per Transaction} & $501,760$ & $321$& $9,302.81$ \\ \hline
\textbf{Per Record} & $31,104$ & $319$ &  $2,617.80$ \\ \hline
\textbf{Fixed Length} & $32$ & $32$ & $32.00$ \\ \hline
\end{tabular}}
\end{table}

\begin{table}[t]
\centering
\caption{The description of fields of audit log JSON packet. The packet has a header and a detail record for each updated property. The detail records can have $(0-n)$ records depending upon the class properties that are being updated.} 
\label{tab:AuditJson}
\scalebox{0.77}{
\begin{tabular}{c|l|l}

\hline
\multicolumn{2}{l|}{\textbf{Field Name}} & \textbf{Description}    \\ \hline 

\multicolumn{2}{l|}{\textbf{AppId}}           & Unique identifier for the application 
                 \\ \hline
\multicolumn{2}{l|}{\textbf{ClassName}}           & The name of updated application class                   \\ \hline
\multicolumn{2}{l|}{\textbf{CreatedDate}}  & Creation date and time for the audit record                         \\ \hline
\multicolumn{2}{l|}{\textbf{EntityId}}   & Unique key for business object  \\ \hline
\multicolumn{2}{l|}{\textbf{EventType}}  & This would be Update, Insert or Delete                        \\ \hline
\multicolumn{2}{l|}{\textbf{Id}}                  & Unique id of the audit record, GUID\\ \hline
\multicolumn{2}{l|}{\textbf{SessionId}}     &  Unique Id for a transaction    \\ \hline
\multicolumn{2}{l|}{\textbf{Url}}           &  Application page creating the audit \\ \hline
\multicolumn{2}{l|}{\textbf{UserId}}           & User id, for the user making the change                       \\ \hline 
 \multirow{4}{1.3cm}{\textbf{Details $(0-n)$}}  
& \textbf{Id}           &    Unique Id for the detail record                     \\ \cline{2-3}
&\textbf{NewValue}           &    Current property value                     \\ \cline{2-3}
&\textbf{OldValue}           &    Old Property value                     \\ \cline{2-3}
&\textbf{Name}           & Name of property e.g owner's name                         \\ \hline
\end{tabular}}
\end{table}

\begin{figure}[t]
\begin{lstlisting}[caption={Blockchain transaction generated after serializing data from the audit log. This transaction is exchanged among the peers during the application runtime.},label={lst:json}, style=json]%[t]

{  "AppId":"USA-FL-0000005",
   "ClassName":"SAGE.BL.InspSystem.PermitInspection",
   "CreatedDate":"\/Date(1532366360155-0400)\/",
   "EntityId":161031,
   "EventType":UPDATE,
   "Id":"9ceb8c2c-154a-49d5-9441-a92600db997b",
   "SessionId":"c66207c8-63be-4703-b858-cbfae98a988e",
   "Url":"\/SAGE\/Building\/Inspection\/InspectionReport.aspx?srcTp=309&srcId=17552018&InspectionTypeId=61663",
   "UserId":666,
   "Details":[  
      { 
         "Id":"fa268eaf-7993-48e3-ae6a-a92600db997b",
         "NewValue":"10",
         "OldValue":"9",
         "PropertyName":"DBVersion"
      },
      {  
         "Id":"ee2cdbc2-9c3a-4bc9-afba-a92600db997b",
         "NewValue":"available after 1:00 pm",
         "OldValue":"available after 2:00 pm",
         "PropertyName":"RequestComments"
      }
   ]
 }
\end{lstlisting}
\end{figure}

\begin{figure}[t]
 \begin{center}
     \includegraphics[width=0.4\textwidth]{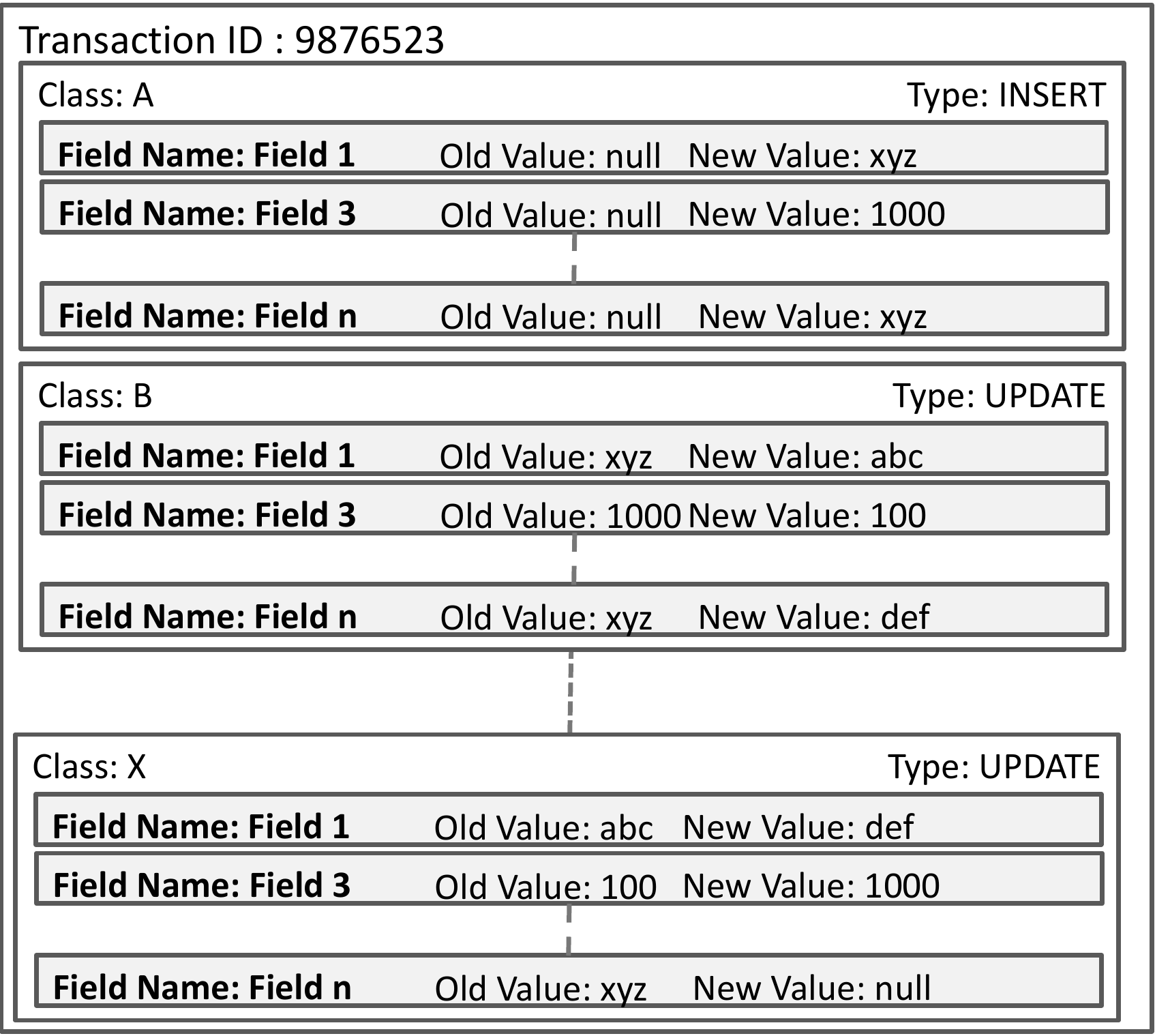}
\caption{Audit generation for a transaction spanning across multiple objects. Object A is inserted, B and X are updated. }
\label{fig:transactionCombine}
 \end{center}
\end{figure}

 \begin{figure}[t]
\begin{center}
    \includegraphics[width=0.35\textwidth]{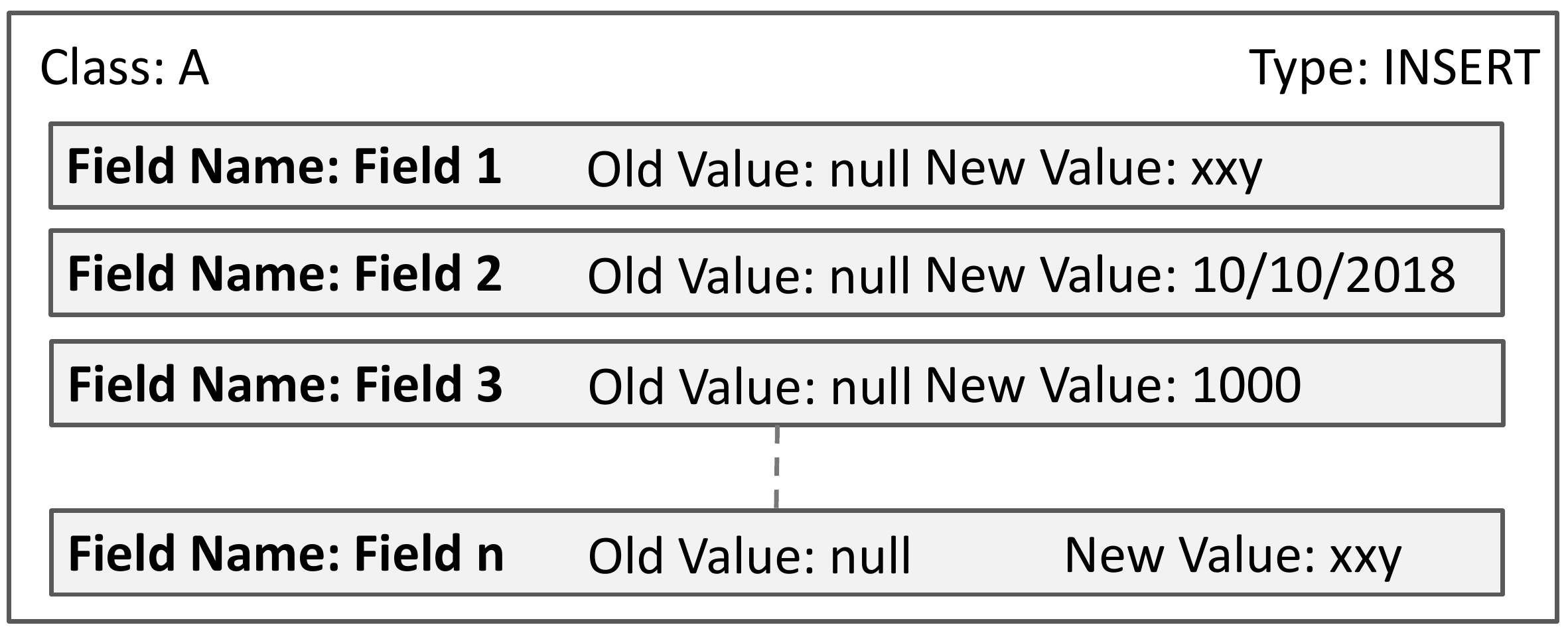}
\caption{Audit Entry generation for a object. Object A is a new object that is being inserted into the database for the first time. An audit log entry only contains one object.}
\label{fig:transPerTable}
\end{center}
\end{figure}

Additionally, each node is required to keep a copy of the blockchain at their servers and maintain a persistent connection with their corresponding application server. Persistent connections are necessary to maintain an up-to-date view of the blockchain in order to process, validate, and forward transactions, as well as to avoid unwanted forks and partitioning attacks that may result from an outdated blockchain view.

\subsubsection{Creating Blockchain Transactions} \label{sec:cbtx}
Once the network architecture is laid out, the next step is to create blockchain-compatible transactions from the audit log data. For that, we convert the audit log data to a JavaScript Object Notation(JSON) format~\cite{rfc4627}. We preferred JSON over other standard data storage formats such as XML, due to its data structure compactness and storage flexibility. To obtain a blockchain transaction, we first pass the audit log data to a function that serializes it to JSON and calls {\em createAudit} REST \cite{Huai2008} web service to create the audit log transaction. Each JSON packet is then treated as a blockchain transaction, and as soon as a node in the network receives a transaction, it broadcasts the packet to the rest of the network. Nodes can connect to multiple peers to avoid the risk of delayed transactions due to malicious peer behavior or network latency. In \autoref{tab:transations}, we show the average transaction size from our sample system for October 2018. In \autoref{tab:AuditJson} we describe the purpose of each field in the audit JSON packet and in \autoref{lst:json}, we show the data structure of the blockchain transaction that is obtained after serializing data from the audit log.  

\BfPara{Log for table/class}
The audit event logger can also create a packet for each object in a transaction. We used this method in the prior work~\cite{AhmadSBM18} and found that the packet size was small, however, the number of web service calls for each application transaction was high. For instance, if a transaction contains 10 classes, it will create 10 web service calls. While 10 calls can be handled by ORM-based audit logs, they are not optimal for blockchain-based audit logs.

\BfPara{Log for transaction}
The audit event logger creates a packet containing all insertions, updates, and deletions, that span across one or more objects, and sends the packet to \bl as shown in \autoref{fig:transactionCombine}. Since the audit log data is consolidated, therefore,  it is hard to search for updates for a specific class, which is a typical use case.  Creating an audit log for a transaction reduces the number of web service calls and improves efficiency, and this design is more suited to blockchain based audit logs.

\subsection{Consensus Protocol} \label{sec:consptcl}
The next phase in the \bl design is the use of a consensus scheme among the peers to develop their agreement over the sequence of transactions and the state of the blockchain. There are various consensus algorithms used in blockchains, such as proof-of-work (PoW), proof-of-stake (PoS), proof-of-knowledge (PoK), Byzantine fault tolerance (BFT), \etc \cite{SaadM18,NguyenK18}. In \autoref{tab:ca}, we compare the popular blockchain consensus algorithms. Notice that PoW and PoS have high scalability and fault tolerance. More specifically, they can scale beyond 10,000 nodes and can tolerate up to 50\% malicious replicas. On the downside, they have low throughput and high confirmation time \cite{CromanDEGJKMSSS16,Vukolic15}. In contrast, PBFT has high throughput and low confirmation time. However, PBFT has low fault tolerance which makes it less suitable for permisionless settings.

For \bl, we use PBFT consensus algorithm~\cite{SukhwaniMCTR17,cachin2016architecture}, which was originally designed to facilitate the decision-making process in a distributed environment. \bl uses a permissioned blockchain system \cite{AndroulakiBBCCC18}, in which all network participants are known to one another, and there is a weaker notion of anonymity. Since our system is primarily a private and permissioned blockchain, therefore, we are not constrained by high scalability challenges. Although in the future, we aim to extend our design to a bigger network, however, at the prototype stage, we are less than 100 peers. Due to high throughput and low latency, naturally, PBFT is more suited for our design. 

\begin{figure}
\begin{center}
    \includegraphics[width=0.45\textwidth]{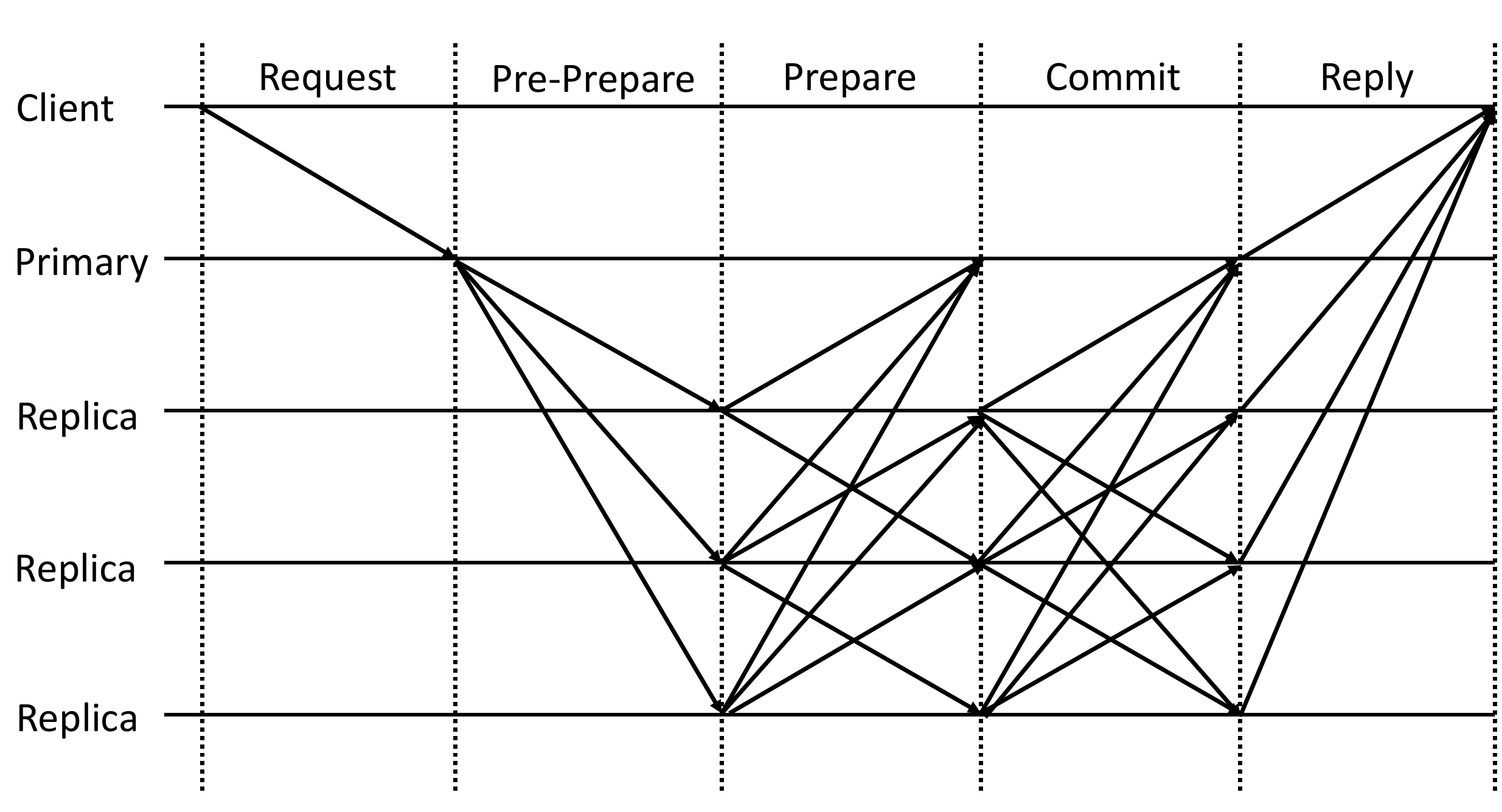}
\caption{An overview of PBFT protocol with client issues a request to the primary replica. The primary then broadcasts the transaction to all the other replicas. The replicas validate the order of the transaction and share their view with each other. Once the client receives the desired number of responses, the transaction is considered validated. The process of transaction verification follows four stages, namely Pre-Prepare, Prepare, Commit, and Reply.}
\label{fig:PBFT}
\end{center}
\end{figure}

\begin{figure}[t]
\begin{center}
    \includegraphics[width=0.43\textwidth]{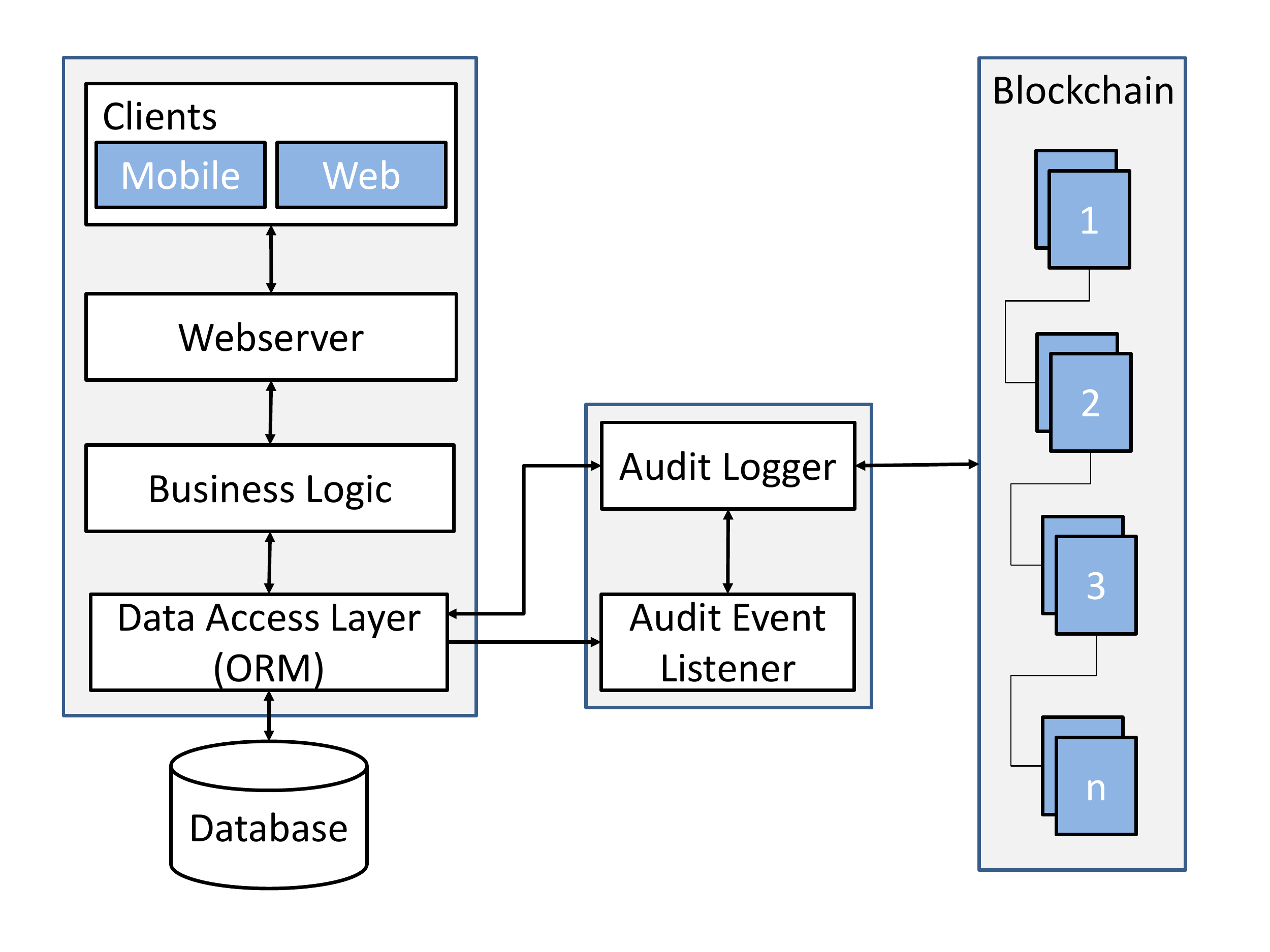}
\caption{Complete system architecture of \bl after blockchain is integrated to the JSON output.}
\label{fig:archi}
\end{center}
\end{figure}

In PBFT, the system comprises of a client that issues a request (transaction), and a group of replicas that execute the request. The primary replica orders transactions and relays them to other replicas. The transaction is processed in four stages, namely pre-prepare, prepare, commit, and reply. When the client receives a minimum of $3f+1$ responses, $f$ being the number of faulty replicas, the transaction processed. In \autoref{fig:PBFT}, we provide an illustration of PBFT, which we later use to design and calibrate \bl. In \autoref{fig:archi}, we show the complete design of \bl, where the blockchain is integrated with the serialized JSON output of the business application. 

\begin{table}[t]
\centering
\caption{An overview of popular consensus algorithms used in blockchains Notice that PBFT has high throughput and low confirmation time.}
\label{tab:ca} 
\scalebox{0.7}{
\begin{tabular}{l|c|c|c}
\textbf{Properties}  & \textbf{PoW}       & \textbf{PoS}       & \textbf{PBFT}        \\ \hline
\textbf{\begin{tabular}[c]{@{}l@{}}Blockchain Type\\ \end{tabular}}         & Permisssionless & Permissionless & Permissioned       \\ \hline
\textbf{\begin{tabular}[c]{@{}l@{}}Participation Cost\\ \end{tabular}} & Yes & Yes  & No    \\ \hline
\textbf{\begin{tabular}[c]{@{}l@{}}Trust Model\\ \end{tabular}} & Untrusted & Untrusted &  Semi-trusted \\ \hline
\textbf{Scalability}                                                         & High               & High                      & Low                            \\ \hline
\textbf{Throughput}                                                          & \textless{}10      & \textless{}1,000 &  \textless{}10,000  \\ \hline
\textbf{\begin{tabular}[c]{@{}l@{}}Byzantine Fault Tolerance\\ \end{tabular}} & 50\%               & 50\%                      & 33\%                            \\ \hline
\textbf{\begin{tabular}[c]{@{}l@{}}Crash Fault Tolerance\\ \end{tabular}}     & 50\%               & 50\%                     & 33\%                           \\ \hline
\textbf{\begin{tabular}[c]{@{}l@{}}Confirmation Time\\\end{tabular}}         & \textgreater{}100s & \textless{}100s & \textless{}10s      \\ \hline

\end{tabular}}
\end{table}

\section{Analysis of \bl}\label{sec:Analysis}
In this section, we analyze various aspects of \bl, including design, complexity, and security analysis.
\subsection{Design Analysis} \label{sec:da}
In \bl, each peer uses the ORM-based audit log application that is connected to a database. Once the ORM observes a change, it updates the database and issues a transaction, and sends it to the primary replica. The primary orders the transaction and broadcasts them to all the other replicas. Upon receiving the transaction, each replica checks if the transaction is valid and follows the correct order. The order of the transaction is ensured by the timestamp, and the ordering rule involves the chronological sequencing of each transaction. In \bl,  the primary preforms transaction sequencing based on the time at which it receives transactions from the application replica. We use this approach as a security design choice to prevent malicious replicas from arbitrarily modifying their transaction timestamps. In the following, we show how transaction sequencing is performed in \bl:
\begin{enumerate}
    \item An application generates a transaction at time $t_{i}$ and the primary receives the transaction at time $t_{j}$.
    \item First, the primary checks if the transaction respects the temporal ordering ($i < j,$ $\forall$ $i,j$ ). This assumption is valid for any real-world system, since each transaction experiences a non-zero delay during transmission.
    \item If the primary observes a violation \ie $i > j$, it assumes that the application replica is misbehaving. Therefore, the primary discards the transaction. 
    \item In the transaction confirmation phase, the active replicas also compare the time at which they receive a transaction to the time of the transaction generation. This serves as an additional security measure to ensure that the policy precedence is respected, even when ignored by the primary.
\end{enumerate}

In \bl, we enforce the ordering of transaction since it is critical in audit log applications. For instance, consider a case in which $tx_{a}$ involves a change made to a class. The next transaction $tx_{b}$ reverses the change made by $tx_{a}$, then it is critical to process $tx_{a}$ before $tx_{b}$. Otherwise, the order will be violated and the audit log will reflect a different state of the database than the actual.

In summary, \bl constitutes of a client (audit log application) that generates blockchain compatible transaction, a primary replica that receives and orders transactions, and a group of active replicas that execute PBFT to generate a blockchain-based audit log. In conventional PBFT, the client is independent of the active replicas that execute the consensus protocol. In \bl, the client is one of the active replicas that issues the transaction. In the verification process, the issuer becomes the client and all other replicas act as validators. 

\BfPara{Key Takeaways} From the design implementation, we had the following takeaways:
\begin{enumerate*}
   \item PBFT-based permissioned blockchains are more suitable for audit log applications.
    \item Extending ORM provides an efficient mechanism of converting database transaction to blockchain compatible transactions.
    \item Existing application can seamlessly integrate with blockchain based audit logs using ORM extension.
    \item REST based web services can also be easily extended to support applications that do not use ORM.
    \item JSON format is the de facto standard for REST API's, and therefore efficient and suitable for an audit log transaction.
\end{enumerate*}

\subsection{Complexity Analysis} \label{sec:complexity}
A key aspect of PBFT-based blockchain systems is the time and space complexity associated with the network and the blockchain size. The time complexity partakes the time taken by replicas to develop consensus on a transaction or a block. The space complexity involves the storage and the search overhead that compounds due to append-only distributed blockchain design. In the following, we analyze these aspects of complexity in \bl.  

\BfPara{Time Complexity}\label{sec:complexity_time2}
To achieve consensus over the state of blockchain with $n$ replicas, $n^{2}-n$ messages are exchanged, as shown in \autoref{fig:PBFT}. Therefore, for each transaction generated within the system, the overall complexity becomes $O(n^2)$. Compared to PoW-based blockchains, in which the consensus complexity is $O(n)$, PBFT has a high message complexity which can lead to system overheads and delays. However, we argue that in PoW-based blockchains systems such as Bitcoin, the total number of active nodes are over 6-8k~\cite{apostolaki2017hijacking}. In comparison, \bl constitutes less than 100 peers. Therefore, it can tolerate this complexity overhead, keeping in view the other benefits associated with PBFT such as high throughput. 

\BfPara{Space Complexity} \label{sec:complexity_time}
The space complexity of the system can be ascribed to the overhead associated with the storage of blockchains at each peer. One major limitation of replacing the client-server model with a peer-to-peer blockchains system is that each peer is required to maintain a copy of blockchain. This leads to a high storage footprint since blockchains are always growing in size. The size footprint also increases the search complexity for transaction verification. For instance, when a newly generated transaction is sent to a group of peers for verification, they validate its authenticity by consulting its history in the blockchain. If the blockchain size is large, the verification time increases. As such, if the rate of the incoming transaction is high, then high verification time may lead to processing overhead, thereby increasing latency and reducing the throughput. In \bl, the space complexity of a system, complementary to any other blockchain system is $O(n)$.

\BfPara{Key Takeaways} From the complexity analysis, we had the following takeaways:
\begin{enumerate*}
    \item PBFT-based based blockchains have high message complexity. Therefore, if the network scales beyond a few hundred nodes, the application may become inefficient. Therefore, we observe a tradeoff between the message complexity and the network scalability.
    \item Generally, the space complexity of blockchain is high, due to the append-only model. In \bl, the space complexity is similar to any other blockchain application.
\end{enumerate*}

\subsection{Security analysis} \label{secLsecAnalysis}
An essential component of our work is the defense against the attacks outlined in the threat model \textsection\ref{sec:thm}. In this section, we discuss how \bl defends against the physical access attack and the remote vulnerability attack.

\BfPara{Physical Access Attack} In the physical access attack if the attacker acquires the credentials of a user, he can make changes to the application data using the application interface. In this case, his activity will be logged in \bl. Since the log is kept in the blockchain by the user, the attacker will not be able to remove the traces of his activity. Therefore, when the attacker's activity is exposed, auditors will be able to track the tampered records and take corrective measures to restore data to the correct state. Moreover, if the attacker is able to get write access to the database, he will be able to change data in different tables. Since the audit log generation is at the ORM level, therefore, these changes will not be present in the audit log. This will enable the auditors to detect malicious activity and take preventive actions.

\BfPara{Remote Vulnerability Attack} In case of a remote vulnerability attack in which the attacker exploits a bug or vulnerability in the application, the audit log will show the effect of the changes or errors resulting from the attack. Additionally, the blockchain will also preserve the tamper-proof state of the audit log prior to the launch of the attack. As a result, the auditor will be able to compare the audit log and the current data to detect changes made during the attack. In the absence of the blockchain, if the attacker corrupts the prior state of the audit log, there is no way auditors can recover from it. However, with \bl, not only the attacks are detected, but the system state is also recovered. Furthermore, for a successful attack in the presence of \bl, the attacker will need to corrupt the blockchain maintained by each node. Based on the design constructs and security guarantees of blockchains, corrupting blockchain repositories of a majority of nodes is costly, and therefore infeasible. 

After realizing that \bl is able to defend against the attacks outlined in our threat model \textsection\ref{sec:thm}, there are however few considerations to be made while using PBFT-based blockchain model. The prior work in this direction does not consider Byzantine behavior among nodes. In \bl, we consider that peers may behave arbitrarily and create confusion in the view of other honest peers. Therefore, we want \bl to be robust against malicious replicas. While other consensus mechanisms such as PoW may withstand up to 50\% of faulty replicas in the system, PBFT, in contrast, has low fault tolerance. In a situation where there are $f$ faulty replicas, a PBFT-based blockchain system needs to have $3f+1$ honest replicas in order to function smoothly. Roughly speaking, PBFT-based blockchains require 70\% nodes to behave honestly in order to avoid disagreements. However, in \bl, we try to raise the threshold of fault tolerance by making minor adjustments to the security design.

\BfPara{Increasing Fault Tolerance} In a situation where there are $r$ honest replicas in a blockchain and the attacker is able to position $f$ faulty replicas such that $4f+1 > f+r$, then the attacker will be able to stop transaction verification and may even cause forks. To counter this, we propose an expected verification time window $W_{t}$ which will be set by the primary replica before passing the transaction to the verifying replicas. The primary replica knows the total number of active replicas in the system and can calculate the total number of messages to be exchanged until the transaction gets verified. In this case, the total number of messages will be in the order of $(f+r)^{2}-(f+r)$. Let $c \times t_{b}$ be the time taken for the transaction confirmation, where $c$ is an arbitrary constant set by the primary replica. Based on these values, the primary replica can set an expected time window $W_{t} \geq c\times t_{b}$ in which it expects all peers to validate the transaction and submit their response. Let $t_{start}$ be the start time at which the primary replica initiates the transaction. If by $W_{t}$ the primary does not receive the expected number of responses from the replicas, it will abort the verification process and notify the auditor.

Depending on the application's sensitivity, the primary replica can either set another optimistic value of $W'_{t}$, where $W'_{t} \geq W_{t}$, and repeat the process or it can simply abort the process and notify the application auditors regarding the malicious activity. We leave that decision to the audit log application and its sensitivity to malicious activities. However, in our experiments, we relax the condition of sensitivity and re-submit the transaction for another round of verification. We set a new expected verification time window $W'_{t}$ and wait for the response. Our choice of relaxing the condition of sensitivity is owing to the unexpected delays in the message propagation; given that our system would run over the Internet. However, if the primary replica does not receive the approval for the second time, it aborts the process and notifies the application.  

\BfPara{Detecting Malicious Nodes} In \bl, we also enable detection of the malicious nodes that corrupt the process of transaction verification. For that, we store the identity of the replica in each iteration of the response. For instance, in the first iteration of $W_{t}$, we note the identity of replicas that send their digitally signed approval for the transaction. Let $h$ be the subset of replicas that send their response in the first iteration, where $h\leq (f+r)$. The primary replica stores the identities of replicas in $h$ and initiates the second iteration at $t'_{start}$ and waits for response till $W'_{t}$. Upon receiving the response in the second iteration, the primary replica updates $h$ and removes the duplicates. By comparing $h$ with the identity of all the replicas, the primary replica can find the malicious replicas and request their removal from the verification process. 

It is possible that an adversary, aware of the two-phased approval process, may attempt to trick the system by sending a response from a subset of malicious peers in each phase of approval. For instance, the adversary can split his set of malicious replicas in $f_{1}$ and $f_{2}$, where $f_{1}+f_{2} = f$. In the first phase of approval, the adversary can send a response from $f_{1}$ replicas. However, the adversary ensures that $ 3f_{1}+1\geq r$, so that the transaction does not get enough approvals to be accepted by the primary replica. The primary replica will append $f_{1}$ to its set of $h$. In the second iteration, the adversary will incorporate signatures from $f_{2}$, and the primary replica will also add them to $h$. As a result, the primary replica will not be able to detect the actual number of malicious replicas in the system. 

\begin{figure*}[ht]
\begin{center}
        \subfigure[Transaction Rate $\lambda= $200 tx/s \label{fig:city}]   {\includegraphics[width=0.31\textwidth]{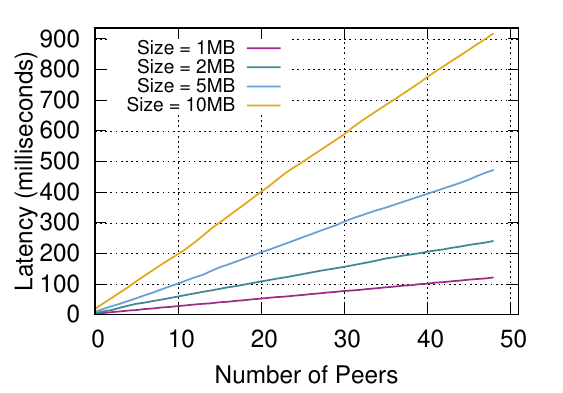}}
	\subfigure[Transaction Rate $\lambda= $3,000 tx/s \label{fig:county}] {\includegraphics[width=0.31\textwidth]{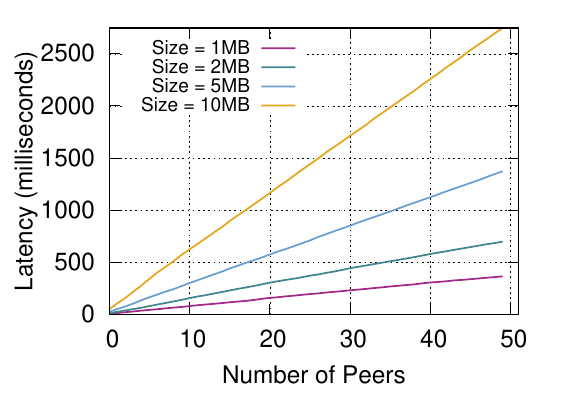}}
	\subfigure[Transaction Rate $\lambda= $6,000 tx/s \label{fig:state}] {\includegraphics[width=0.31\textwidth]{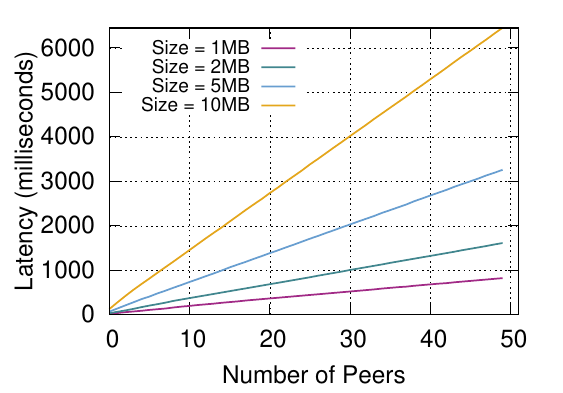}}
	
\caption{ Time taken to reach consensus at different types of audit transaction with varying transaction rate $\lambda$ (200-6,000 tx/second). Notice that as the network size and the payload size increases, the confirmation time for a transaction increases. Also, it can be seen that the as $\lambda$ increases, confirmation time increases. Naturally, this can be associated with high verification delays with the bulk of the incoming transactions. }\label{fig:R1}
\end{center}
\end{figure*}

To counter that, we randomize the two-phase approval process to $v$-phase approval process, $v$ may take any value of the primary replica's choice. When the transaction fails the first attempt, the primary replica can either abort or continue the approval process. Continuing from the above-outlined scenario, if $v=3$, then the attacker will either have to include one of $f_{1}$ or $f_{2}$ replicas in the third phase. And if the primary replica iterates one more time, the adversary will be bounded to include the set of replicas that he did not include in the previous iteration. As such, the primary replica will notice the incoherence in the response of a few replicas in each iteration of the approval process, and the adversary will risk exposing his malicious replicas. Although this procedure ensures high security and the ultimate exposure of adversary in the process of verification, it is, however, time-consuming and may lead to a transaction stall. Again, we leave this to the primary replica, which can make decisions that suit the application requirements.

\BfPara{Key Takeaways} From the security analysis, we had the following takeaways:
\begin{enumerate*}
    \item \bl counters the conventional audit log attacks namely the physical access attack and the remote vulnerability attack.
    \item Additionally, \bl also makes audit logs secure against Byzantine behavior, tolerating up to 30\% malicious replicas.
    \item Leveraging the design policies in permission settings, \bl is able to detect malicious replicas. 
\end{enumerate*}

\section{Experiment and Evaluation } \label{sec:eval}
In this section, we present experiments carried out to evaluate the performance of \bl. First, we extended the {\em nHibernate} ORM to generate a serialized JSON output in the form of transactions as shown in \autoref{lst:json}. The transactions are broadcast to the network where a \bl blockchain instance is configured at each node. For experiments, we used sockets to set up the network and a NodeJS client to receive JSON transactions. 

\BfPara{Simulation Environment}
We simulated our blockchain network using a LAN setup at our research lab. We used 20 machines, each running the Linux OS with Intel Core i5 processor and a 16MB RAM. Next, we set up a virtual environment at each node to construct a multi-host network. We assigned port numbers and sockets to each host that acted as a peer. The socket connections were used to exchange data with peers using IP addresses and port numbers. Each peer was equipped with a JSON master list that contained the information of all the other nodes. Data packets of the desired size were generated and broadcast over the network. We encoded the PBFT protocol in NodeJS and executed it over all the peers. The selection of the primary replica can be done using any method suitable for the application. In \bl, in each iteration, we selected the primary in Round-robin manner. To reflect the real-world delays in our simulation, we manually added a round-trip delay of 100ms in each transaction broadcast over the network. Finally, once the transaction obtained sufficient approvals, it was added to the blockchain of the primary replica, and subsequently, all the other replicas. 

We evaluate the performance of our system by measuring the latency over the consensus achieved by peers. We increase the transaction payload size from 2MB to 20MB and the rate of transaction $\lambda$ from 200 transactions per second to 6,000 transactions per second. By adjusting these parameters, we monitor the time taken by peers to approve the transaction. Let $t_{g}$ be the transaction generation time, and $t_{c}$ be the time at which it gets approval from all active peers. In that case, the latency $l_{t}$ is calculated as the difference between $t_{c}$ and $t_{g}$ ($l_{t} = t_{c} - t_{c}$,  where $t_{c} > t_{g}$). We report the simulation results in \autoref{fig:R1}. 

\BfPara{Simulation Results}Our results show that irrespective of the payload size, the latency margins remain negligible as long as the number of peers is less than 30. As the size of the network grows beyond 30 nodes, the latency factor increases considerably. Furthermore, we also notice that a sharp increase in latency when the payload size changes from 5--10MB and a negligible change in latency when the payload size changes from 15--20MB.

We also noticed that as the rate of transaction $\lambda$ increases from 200 transactions per second to 6,000 transactions per second, the confirmation time for transaction also increases. Intuitively, this can be attributed to the processing overhead caused by the increasing rate of $\lambda$ at each replica. However, it can be observed from \autoref{fig:state} that within a network size of 50 peers, \bl has the capability of processing 1,000 transactions per second, with the payload size of 10 MB. This payload size is equivalent to 10 blocks in Bitcoin. For the payload size of 1MB, \bl achieves a throughput of 6,000 transactions per second. Considering low throughput of conventional blockchains (3--7 transactions/second in Bitcoin), \bl achieves high throughput. This also justifies our choice of using PBFT as consensus scheme for our system.  

Evaluation parameters obtained from our experiments can be used to define the block size and the network size, specific to the needs of the application. As part of our future work, we will use these parameters along with other consensus schemes to find optimum block size and the average block time for the audit log application. By varying consensus schemes, we will be able to compare and contrast the performance of various design choices and select the best that can be used for \bl.

\section{Discussion and Future Work} \label{sec:discusion}
With \bl, we were able to meet our overall objective of securing audit logs using blockchains. We show with theoretical analysis and simulations that our system is secure and efficient, and it achieves high throughput(\textsection\ref{sec:eval}) by using the PBFT consensus protocol. In \bl, audit log transactions were seamlessly generated with minor changes to the existing system. Moreover, \bl can be plugged into any enterprise business application, that consumes a REST API to send audit log data as a transaction. In summary, we successfully extended our application into the blockchain paradigm to harden its security and increases the overall trust in the application. Our system is robust against the physical access attack and the remote vulnerability attack.

\BfPara{Limitations}
Despite all the promising outcomes, there are, however, two major limitations in \bl. The first constraint is the high message complexity due to PBFT, and the second is a high storage footprint due to data redundancy in the blockchain design. Since in PBFT, the message complexity is high ($O(n^2)$), therefore, in adverse network conditions, PBFT may perform poorly, compared to other consensus protocol~\cite{AbrahamM17}. In spite of these limitations \bl performs within the requirements of our application, and could support PayPal \cite{gobel2017increased} which processes 170 transactions/second, however, our solution would not be feasible for Visa which has a transaction rate of 2000 transactions/second \cite{croman2016scaling}. Secondly, audit logs by design have a high storage footprint, as each transaction in the system has a corresponding entry in the audit logs. In \bl, the problem is further increased since transactions are replicated on multiple peers, resulting in high storage overhead. 

Keeping in view these limitations, we propose that high message complexity can be resolved by using other newly proposed consensus algorithms such as Clique \cite{Angelis18}, that belongs to the family of Proof-of-Authority consensus protocols. Clique has a message complexity of $O(n)$, which is considerably lower than PBFT and PoW. Using Clique may allow us to support a larger number of peers, achieve high throughput, and reduce confirmation delays of transactions in \bl. However, in Clique, peers run into the risk of multiple views at the same time. In blockchains, this inconsistency is called a blockchain fork. These forks can lead to temporary or permanent partitioning in the network. Currently, we are exploring methods of fork resolution in Clique, and therefore applying it in \bl is part of our future work. 

The space complexity can be reduced by adding data retention policy and purging data after its fixed retention time. This would optimize the overall size of the blockchain, and lead to less storage and search complexity. In addition to these two schemes, we also propose two other optimization strategies to meet the design limitations in \bl. 

Another limitation in \bl is the weak link between the application and the audit log. In the current implementation, if the application itself is compromised, and subsequently the audit log generation fails, then \bl will not be able to detect the fault at the application. At present, \bl enables applications to seamlessly integrate with blockchain system and benefit from it. Therefore, \bl remains agnostic to the application itself and the data being produced by it. As a result, we observe a trade off between the seamless integration of audit logs with the application and the enhanced security of the audit log generation interface. Currently, \bl is designed to facilitate the integration of audit logs with eGovernment application. In future, we also aim to focus on detection application-level faults in \bl. 

The latency is a critical problem in distributed systems, which can be 1) latency due to the consensus scheme operation, and 2) latency due to network conditions.  To minimize latency due to consensus, we select consensus algorithms, such as PBFT, which is known to provide low latency and high throughput compared to other popular schemes such as PoW.  We note that such a choice comes at a certain cost: PoW is known to have better security, since it tolerates up to 50\% Byzantine nodes while PBFT tolerates only 30\% \cite{BanoSBAMMD17}. Acknowledging that, and giving latency a higher priority over security, in \bl, we made the consensus choice to minimize the latency. 

The other component of latency is due to the network, which includes transmission and propagation delays under a certain payload size. In \bl, and as shown in \autoref{fig:R1}, with a payload of 10MB and a network of 50 replicas, the transaction confirmation experiences a delay of 6 seconds. In \bl, this is an upper bound on the end-to-end latency, which is considerably low compared to 600 seconds of delay in Bitcoin. For our Enterprise application, this delay is tolerable. However, if \bl is to be extended for applications with larger payloads, we suggest two improvements as the latency increases. First, the communication medium between applications can be enhanced to support high bandwidth. Second, localities could be exploited to host applications within the same autonomous system to reduce propagation delays. Implementing these improvements is a future work.

 \begin{figure}[t]
\begin{center}
    \includegraphics[width=0.45\textwidth]{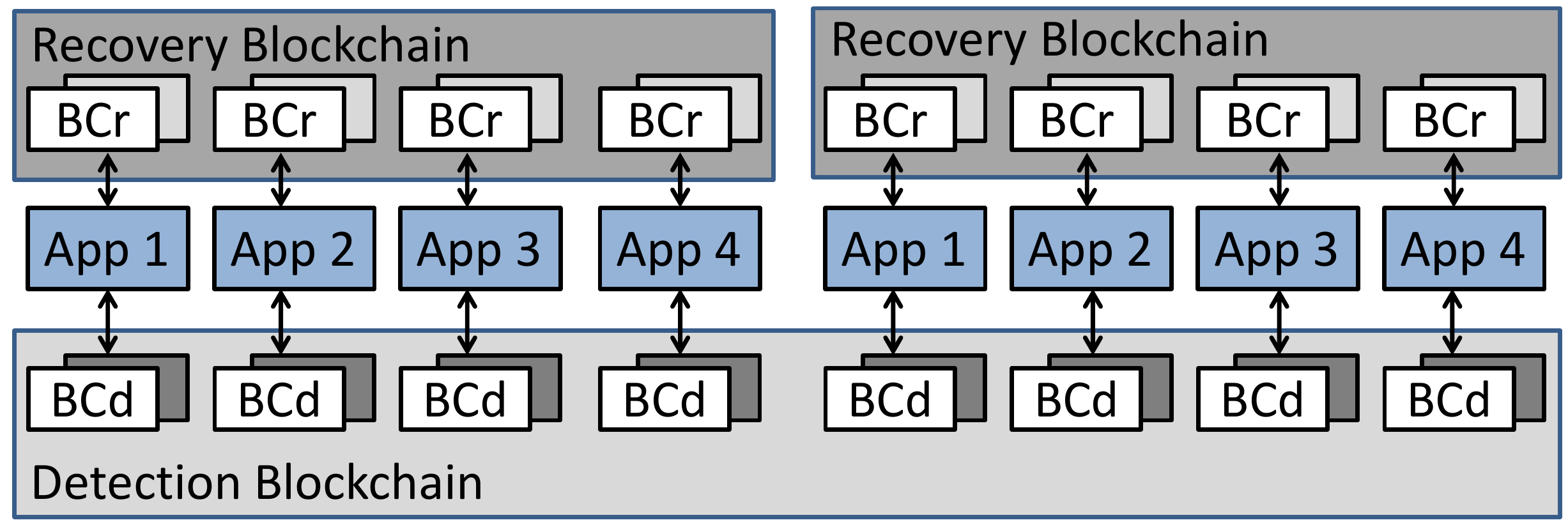}
\caption{Audit log block chain detection vs recovery, there are Recovery blockchain and detection blockchain, the recovery blockchain has less nodes and stored complete log, and the detection blockchain stores a hash value of block and has more nodes.}
\label{fig:DesignDetVsRec}
\end{center}
\end{figure}

\BfPara{Optimization}To increase the performance and to keep the audit log tamper-proof, we propose having two sets of blockchains, namely the {\em recovery blockchain}, and the {\em detection blockchain}. In \autoref{fig:DesignDetVsRec} we provide a system overview of this two blockchain system. The recovery blockchain stores the complete audit log transaction, including details of all data changes in an application-level transaction. The {\em recovery blockchain} can be used to restore data to its prior state, which would be the state of data before an attack. The {\em recovery blockchain} would require more space, and longer consensus time due to large transaction audit data packets. The number of peers $k$ in the {\em recovery blockchain} can be kept small to ensure immediate consensus and avoid delays. Since the security of PBFT relies on the faithful execution of the protocol by at least 70\% replicas, therefore for a baseline, the minimum size of the {\em recovery blockchain} must be four nodes considering one malicious replica ($k\geq4$).

The {\em detection blockchain} can be used to detect audit log tampering only. It will not have the information to recover the audit log to a correct state before the attack. The business application will generate a cryptographic hash using the audit transaction. The hash and a unique transaction identifier will be stored in the blockchain. In the case of data tampering in the audit log, the newly computed hash will not match the hash stored in an audit log. This will indicate that the audit log has tampered. Once tampering is detected, application administrators could use corrective measures to fix the security breach. Atop that, data can could be restored to the previous state by using database backups. The recovery blockchain which will have $K_d$ peers, where $K_d > 2K$. Therefore, the adversary will have to compromise twice as many nodes to tamper the system without being detected. This optimization increases security and provides a second layer of defense.

Despite the existing challenges, \bl is a feasible approach towards blockchain-based secure audit logs. Extending the capabilities of the prior work, \bl brings the theoretical foundations into practice and as shown in \autoref{sec:eval}, it has been deployed and instrumented in a real blockchain network. Moreover, \bl is also capable of ensuring operational consistency even in the presence of Byzantine replicas. Therefore, it is
a better candidate for the audit log security and can be applied to eGovernment solutions. 

\section{Conclusion} \label{sec:conclusion}
In this paper, we present a blockchain-based audit log system called \bl, that leverages the security features of blockchain technology to create distributed, append-only, and tamper-proof audit logs. We highlight the security vulnerabilities in existing audit log applications and propose a new design that extends NHibernate ORM to create blockchain-driven audit logs. For our experiment, we used an application provided by ClearVillage inc to generate transactions from audit logs, and record them in our custom built blockchain. By design, \bl is agile, plug and play, and secure against internal and external attacks. In the future, we will extend the capabilities of \bl by deploying it in a production environment and explore various performance bottlenecks and optimization techniques.

\BfPara{Acknowledgement} This work was supported in part by KRF grant number NRF-2016K1A1A2912757 and by the Air Force Material Command award FA8750-16-0301.

\bibliographystyle{IEEEtran}
\bibliography{references,conf}

\end{document}